\documentclass[a4paper, english, 12pt, reqno, dvipsnames]{article}

\usepackage{amsmath}
\usepackage{a4wide}

\usepackage[english]{babel}
\usepackage[utf8]{inputenc}
\usepackage{csquotes}
\usepackage{amssymb}
\usepackage{tikz,tikzscale}
\tikzset{>=latex}
\usepackage{booktabs,multirow}
\usepackage{amsfonts}
\usepackage{bm}
\usepackage{subcaption}
\usepackage[countmax]{subfloat}
\usepackage{algorithm,algorithmic}

\newcommand{\corr}[1]{#1}

\newtheorem{theorem}{Theorem}[section]

\numberwithin{equation}{section}

\usepackage{xcolor}
\definecolor{goldenyellow}{rgb}{1.0, 0.87, 0.0}
\newcommand\circlearound[1]{\scalebox{0.8}{\tikz[baseline]\node[circle,inner sep=1pt,draw,anchor=base] {\textup{#1}};}}

\usepackage[colorlinks = true, linkcolor = blue, citecolor = blue, urlcolor = blue]{hyperref}
\hypersetup{final}

\title{Parameter estimation for cellular automata}

\author{
\begin{minipage}{\textwidth}
\small \textbf{Alexey Kazarnikov}, Interdisciplinary Center for Scientific Computing (IWR), Heidelberg University, Mathematikon, Im Neuenheimer Feld 205, 69120 Heidelberg, Germany,
kazarnikov@gmail.com
\end{minipage}
\and
\begin{minipage}{\textwidth}
\small \textbf{Nadja Ray}, Mathematical Institute for Machine Learning and Data Science, Catholic University of Eichstätt-Ingolstadt, \corr{Hohe-Schul-Str. 5}, 85049 Ingolstadt, Germany, \corr{nadja.ray@ku.de}
\end{minipage}
\and
\begin{minipage}{\textwidth}
\small \textbf{Heikki Haario}, School of Engineering Science, Lappeenranta--Lahti University of Technology, P.O. Box 20, 53851 Lappeenranta, Finland, heikki.haario@lut.fi
\end{minipage}
\and
\begin{minipage}{\textwidth}
\small \textbf{Joona Lappalainen}, School of Engineering Science, Lappeenranta--Lahti University of Technology, P.O. Box 20, 53851 Lappeenranta, Finland, joona.lappalainen@lut.fi
\end{minipage}
\and
\begin{minipage}{\textwidth}
\small \textbf{Andreas Rupp} (corresponding author), \corr{Department of Mathematics, Faculty of Mathematics and Computer Science, Saarland University, DE-66123 Saarbrücken, rupp@math.uni-sb.de}
\end{minipage}}

\date{
\begin{minipage}{.9\textwidth}
\small\textbf{Acknowledgements.} We kindly acknowledge the fruitful discussions with Alexander Prechtel and Simon Zech on cellular automaton methods in soil science.\\
A.\ Rupp has been supported by the \corr{Research Council} of Finland's \corr{decision} numbers 350101, 354489, 359633, 358944, and Business Finland's project number 539/31/2023. This research has also been supported the German research foundation's research unit 2179 \emph{MAD Soil}, and the DAAD PPP Finnland under grant number 57610378.
\end{minipage}
}

\begin{document}

\maketitle
% \markleft{\textsc{A.s Kazarnikov, N. Ray, H. Haario, J. Lappalainen, and A. Rupp}}

\begin{abstract}
 \noindent Self-organizing complex systems can be modeled using cellular automaton models. However, the parametrization of these models is crucial and significantly determines the resulting structural pattern. In this research, we introduce and successfully apply a sound statistical method to estimate these parameters. The decisive difference to earlier applications of such approaches is that, in our case, both the CA rules and the resulting patterns are discrete. The method is based on constructing Gaussian likelihoods using characteristics of the structures, such as the mean particle size. We show that our approach is robust for the method parameters, domain size of patterns, or CA iterations.
 \\[1ex] \noindent \textsc{Keywords.}
 Cellular automaton, discrete model, parameter identification, statistical approach.
\end{abstract}

% \noindent\textbf{This article uses empirical cumulative distribution functions, previously used to estimate parameters in continuous models (parameters in partial differential equations), to estimate parameters in completely discrete models. To this end, we demonstrate that our method can reliably estimate the parameters of a cellular automaton method describing self-organizing structures.}

\section{Introduction}
Cellular automaton (CA) models are widely used to describe self-organizing, complex systems such as tumor growth~\cite{moreira2002cellular}, protein bioinformatics~\cite{xiao2011cellular}, chemical reactions~\cite{menshutina2020cellular}, formation and turnover of soil microaggregates~\cite{RayRP17,Zech2022}, geospatial environmental modeling~\cite{ghosh2017application}, urban planning~\cite{sante2010cellular}, crowd evacuation \cite{yang2011review}, traffic flow~\cite{tian2021review}, and microstructure evolution in metal forming~\cite{yang2011review}. Within the framework of a CA, distinct states are assigned to so-called cells. These states may change according to prescribed transition rules depending on the states of the neighboring cells (e.g., within the %von Neumann 
neighborhood %(VNN)
of a specific size).

Typically, several parameters influence a CA's rules, significantly determining the patterns the CA produces. \corr{Thus, it is crucial to determine these parameters since CAs produce significantly flawed simulations if they are set incorrectly. Exemplary, if we try to model the growth of microaggregates in soil, wrong parameters can lead to very large aggregates in computer simulations. Still, these large aggregates cannot be reproduced in \emph{in vitro} experiments. If we model cancer development with CAs \cite{cooper2020}, wrong parameters can produce inaccurate growth and decay rates.}

% One prototype of such a parameter is the size of the VNN.
However, reasonable parameter choices are often hard to identify. An apparent reason is the inherent randomness of CAs that lead\corr{s} to stochastic cost functions in the parameter identification scheme. While the literature on cellular automaton applications is huge, the literature on parameter calibration of CA models is much more sparse.  To calibrate a model in urban dynamics, i.e., the spread of cities, parameter estimation is conducted via a genetic algorithm in ~\cite{Li2007genetic}. In this case, the transition rules depend on geographical variables, physical constraints, and uncertainty. Knowledge about the parameters can be used to improve urban planning towards compact cities, e.g., concerning energy and sustainable land usage. Likewise, neural networks were used to predict parameters in urban planning~\cite{Yeh2004} based on satellite remote sensing data and GIS (Geographic Information System). Finally, parameter estimation for CAs for predicting wildfire in Africa is found in~\cite{couce2010statistical}. The fire propagation was assumed to depend on environmental (vegetation, fuel/litter load, wind) and climatic factors. \corr{Here, parameter estimation was implemented by minimizing the KL (Kulback-Leibler) distance between modeled and observed fire extension histograms.}

%These factors were multiplied and scaled with a parameter $k$, estimated by a statistical evaluation of satellite data of wildfires. More precisely, the choice of $k$ maximizes the agreement between modeled and observed fire extension histograms. The agreement was formulated to minimize the KL (Kulback-Leibler) distance between the histograms. 

A common issue in the CA model\corr{'s} parameter estimation approaches is the lack of statistics. A more or less ad-hoc cost function is formulated and optimized, but no uncertainty quantification is presented.  In this research, we develop a sound statistical approach to the CA parameter estimation problem. The starting point is an analogy of CA patterns with Turing models. \corr{These continuous models were proposed by Alan Turing in the seminal work \cite{Turing1952} as a hypothetical mechanism describing the symmetry-breaking phenomenon at the early stage of morphogenesis. They later found applications in different areas, including modeling chemical reactions \cite{lengyel1991,lengyel1992}, describing population dynamics \cite{zhu2024}, social interactions \cite{ke2022,yuan2023,zhu2023,Li2024}, and even morphochemical processes \cite{bozzini2015,sgura2019,frittelli2024}. For both Turing models and cellular automata, randomized initial values lead to random patterns.} Our \corr{proposed statistical} approach was earlier used to identify parameters of Turing models in continuous and network domains \cite{KazarnikovH20,Zhu2022,Zhu2022a}, and initially introduced to calibrate chaotic dynamical systems
\cite{haario2015, springer2019}.
It is designed for systems with stochastic outcomes that may be due to unknown, randomized initial conditions or stochasticity of the model itself. The approach is based on creating statistics for scalar-valued characteristics computed \corr{directly from} the patterns. The decisive difference to the earlier applications in \cite{KazarnikovH20,KazarnikovH22} is that in this work, both the CA rules and the resulting CA patterns are discrete, while the algorithm of \cite{KazarnikovH20} has been developed for and applied to partial differential equation (PDE) based models, which yield continuous functions as solutions, that depend continuously on the model parameters. Thus, the previous algorithm took advantage of these facts, e.g., by using \corr{different Lebesgue and Sobolev norms} to characterize properties of the patterns produced by the forward model. As opposed to this, CA models yield completely discrete and discontinuous results. If these results are interpreted as functions, these functions only take values in discrete subsets, such as in $\{0,1\}$, prohibiting the evaluation of \corr{spatial gradients}.

Consequently, their discrete analogs must replace the norms/metrics on continuous function spaces. On the other hand, various measures such as the Minkowski characteristics \cite{ArmstrongMRLASS19} or the mean particle size/particle size distribution are widely used to characterize structures. Our statistical approach can directly employ those measures. Indeed, considering such measures leads to sharper estimates than those adopted from the continuous model norms.

\corr{The main novelties of our algorithm comprise the ability to use arbitrary quantities in the parameter estimation process. Thus, we can build our parameter estimation process on relevant quantities in the respective scientific field. In soil science, such quantities are, e.g., the total surface, number of particles, average particle size, compactness ratio, and particle size distribution (which is not a scalar); see \cite{RuppGMPRRT19} for a detailed overview. We illustrate this feature using the average particle size and particle size distribution.}

\corr{
As the CA simulation is inherently stochastic, standard likelihood constructs are not available, and we here indeed deal with a situation often discussed under the title 'intractable likelihood.'  The statistical analysis for such cases is typically carried out with  'likelihood-free inference' methods, among which the ABC (Approximate Bayesian Computation) approach is the most common. Our approach's main difference and novelty is that we create a likelihood, even an empirically Gaussian one, by considering eCDF vectors derived from data. For scalar-valued data discussed above, the eCDF vectors can be directly computed. Still, we need to map to scalars for the 2D patterns created by CA simulations before eCDF vectors can be computed. Here, we use the $L^1$ distance between pairs of patterns. The particle size distribution is a nice feature, directly providing an eCDF vector. The concatenations of Gaussian vectors are again Gaussian, and the joint mean and covariance can be readily numerically estimated.}

The paper is structured as follows: In Section~\ref{SEC:CAM}, we introduce our cellular automaton method, which is used as \enquote{forward model} to produce patterns for our parameter estimation method. The parameter estimation method is outlined in detail in Section~\ref{SEC:Parameter_estimation}. In Section~\ref{SEC:Results}, we discuss the results of our parameter estimation method. This includes a sensitivity analysis concerning the parameters of the CA and the parameter estimation method. We conclude the paper with an outlook \corr{to} future research.

Notably, all used software for this paper is made publicly available in two software projects:
\begin{itemize}
 \item The implementation of the CA model can be found in \cite{RuppZL22}. It is performed in C++ with a MATLAB interface using the mex compiler and a Python interface using the just-in-time compilation provided by HyperHDG \cite{RuppGK22,HyperHDGgithub}, which builds on Cython\footnote{\url{https://cython.org/}}.
 \item The package for parameter estimation can be found in \cite{RuppHK22}. It contains a Python implementation of the presented empirical cumulative distribution function (eCDF) based approach. This package can also be obtained from PyPI\footnote{\url{https://pypi.org/project/ecdf-estimator/}}.
\end{itemize}
\section{Cellular automaton method}\label{SEC:CAM}
We now describe the cellular automaton method and illustrate it in two spatial dimensions, although it is implemented to work accordingly in any positive-integer dimensional setting. The complete, C++-based implementation can be found in \cite{RuppZL22}.

\subsection{Setting of CA model}
The CA model consists of a discretized domain, typically a $d$-dimensional cube, comprising $N^d$ non-overlapping small cubes (so-called \emph{cells}) $c_m$, $m=1,\dots,N^d$, each having identical volumes. Within this domain, the spatiotemporal distribution of two phases \circlearound{0} (e.g., \emph{void}, white in Figure \ref{FIG:move}) and \circlearound{1} (e.g., \emph{solid}, black in Figure \ref{FIG:move}) is considered. We write $c_m = 0$ if cell $c_m$ attains state \circlearound{0} and $c_m = 1$ if $c_m$ attains \circlearound{1}. At the initial time, to all cells, either of the values \circlearound{0} or \circlearound{1} is assigned, e.g., randomly. After that, the cells are redistributed within the domain in every time step according to specific parameter-dependent jumping rules, see Section \ref{SEC:Jumping_rules}. This results in the two-phase system's temporal evolution (self-organization) and a final arrangement of the cells (pattern).

In this study, we prescribe the porosity
\begin{equation*}
 \theta = \frac{\sum_{m = 1}^{N^d} (1 - c_m)}{N^d} = \frac{\textup{number of cells with state \circlearound{0}}}{\textup{number of cells}} \in (0,1)
\end{equation*}
of the system and derive the number of cells of type \circlearound{1}. These are then randomly distributed in the cubic domain $N^d$ at initial time $t_0=0$. The remaining cells are associated with phase \circlearound{0}, i.e., the initial state $s_0$ is an element of the pattern space $P=\{0,1\}^{N^d}$. In fact, $s_0$ is generated by randomly selecting \corr{$N^d \theta$} cells out of the $N^d$ cells (simple random sampling without replacement), i.e., $s_0 \sim U^{\theta}\{0,1\}$. Moreover, we assume the domain to be periodic, i.e., we identify the left and the right boundary and the top and the bottom boundary with each other and likewise in higher dimensions.

\subsection{Jumping rules for the CA model and related parameters}\label{SEC:Jumping_rules}
The jumping rules of our CA are designed \corr{such} that phase \circlearound{1} is compacted; see Figure \ref{FIG:move} for an illustration. The jumping rules depend on the choice of the neighborhood (see Figure~\ref{FIG:vnn}), which in turn depends on the jump parameter~$\sigma$ in some parameter set~$\Pi\subset\mathbb{N}$, and the evaluation of the attractivity of new spots. Thereby, the parameter choice highly influences the self-organization of the system and the pattern obtained; see also illustration in Figure~\ref{fig:domain_porosity_jumpParameter} and~\ref{fig:time_porosity_jumpParameter}.
\begin{figure}
 \includegraphics[width=\textwidth]{pictures/movement}
 \caption{The application of CA jumping rules within one time-step for the case $\sigma = 2$: First, all agglomerates move in random order to the positions with the highest number of neighbors: \protect\circlearound{B} changes its position, but \protect\circlearound{C} remains in its place because every possible movement decreases the number of neighbors (left picture). Then, all single cells (part of agglomerates or not) move to maximize the number of their neighbors. Arrows indicate the latter movement (center picture). The right picture shows the resulting structure.}\label{FIG:move}
\end{figure}

\subsubsection{Von-Neumann neighborhoods}
First, the parameter $\sigma \in\Pi$ determines the size of the neighborhood, which is considered to decide about possible jumps of single cells with state \circlearound{1} to more attractive spots. It describes the range of the von Neumann neighborhood (VNN), which is the most commonly used distance in CA applications, given by  
\begin{equation*}
\operatorname{range} \text{VNN}(\text{cell}) = \max\{1,\sigma\}.
\end{equation*}

As illustrated in Figure \ref{FIG:vnn}, for a single cell \circlearound{A}, the VNN of size $1$ ($\sigma=1$) consists of \circlearound{A} and its face-wise neighbors, i.e., four neighbors in the two-dimensional space (illustrated in black in Figure~\ref{FIG:vnn}). A VNN of size $2$ ($\sigma=2$) consists of the VNN of size $1$ and all face-wise neighbors of all cells contained in the VNN of size $1$, i.e., 12 neighbors in the two-dimensional space (illustrated in black and red in Figure~\ref{FIG:vnn}), etc. Depending on the choice of the parameter $\sigma$, the single cells of type \circlearound{1} can move within a smaller or larger region to find more attractive spots; see Section \ref{SEC:Movement} below.

\corr{More formally, the VNN of range $r$ around cell $c$ consists of all cells that can be reached by $c$ when it performs at most $r$ consecutive moves into one of its face-wise neighbors. Analogously, the VNN of range $r$ of a set of cells $C$ comprises all cells that can be reached by any cell in $C$ when it conducts at most $r$ consecutive moves into its face-wise neighbors.}

Second, the parameter $\sigma$ is used to determine the size of the VNN, in which agglomerates (ag), i.e., composites of face-wise connected cells of type \circlearound{1}, are allowed to move. The following definition realizes this:
\begin{equation*}
 \operatorname{range} \text{VNN}(\text{ag}) = \max\left\{1,  \left\lfloor \frac{\sigma}{\sqrt[d]{\mu(\text{ag})}} \right\rfloor \right\},
\end{equation*}
where $\lfloor\cdot\rfloor$ indicates the floor function, i.e., rounding down to the next integer, $d$ is the spatial dimension (e.g., two), and $\mu(\text{ag}) > 1$ is the size of ag. It is defined as the number of cells of which ag consists. In the case of $\circlearound{B}$ or \circlearound{C} in Figure \ref{FIG:move}, for instance $\mu(\circlearound{B}) = \mu(\circlearound{C}) = 4$ holds.

\corr{From the above equations, it is obvious that $\sigma$ is the decisive parameter when it comes to the question of how far a cell or aggregate can move and what the patterns that the CA produces look like, see Figure \ref{fig:domain_porosity_jumpParameter}. The equations themselves stem from the reasoning that soil aggregates diffusion is proportional to the inverse of their diameter, which is $\sim 1$ for single cells, and $\sim \sqrt[d]{\mu(\text{ag})}$ for a ball-like aggregate.}

\subsubsection{Movement of agglomerates and single cells}\label{SEC:Movement}
\begin{figure}[t]\centering
 \includegraphics[width=.5\textwidth]{pictures/stencil}\caption{Illustration of VNN around \protect\circlearound{A} of range 1 (black), 2 (red), and 3 (yellow) in two spatial dimensions.}\label{FIG:vnn}
\end{figure}
\begin{algorithm}
 \begin{algorithmic}[1]
 \STATE \textbf{Input}: $s \in P$, the current state of the CA model
   \STATE \textbf{Input}: $\sigma \in \Pi$, the jumping rate parameter
 \STATE Construct set of agglomerates $A_s$ in state $s$.
   \FOR{each agglomerate $\textup{ag} \in A_s$ (in random order)}
    \STATE Calculate VNN(ag) depending on jump parameter $\sigma$
    \STATE Evaluate the attractivity of all possible positions in VNN(ag)
    \STATE Choose a new position for $ag$, maximizing the attractivity 
    \STATE Move ag to the new position
   \ENDFOR
    \STATE Construct the set of all cells with state \circlearound{1} $C_s$.
   \FOR{each cell $c \in C_s$ (in random order)}
    \STATE Calculate VNN(c) depending on jump parameter $\sigma$
    \STATE Evaluate the attractivity of the members of the VNN(c)
    \STATE Choose a new position for $c$, maximizing the attractivity
    \STATE Move cell $c$ to the new position
   \ENDFOR
  %\ENDFOR
  \STATE \textbf{Output}: Updated state vector $s$, i.e., $F_\textup{CA}(\sigma)s$
 \end{algorithmic}
 \caption{Cellular automaton, the definition of non-linear forward operator $F_\textup{CA}(\sigma) \colon P \rightarrow P$\corr{, where $P=\{0,1\}^{N^d}$ denotes the pattern space}. This operator implements a single time-step of the CA model. In this work, we define as a pattern the state vector $s \in P$ obtained by applying operator $F_\textup{CA}(\sigma)$ ${n^{*}}$ times to the initial random state $s_0 \in P$, i.e. $s = [F_\textup{CA}(\sigma)]^{n^{*}}s_0$.}\label{algo:each_iter}
\end{algorithm}
Starting from an initial state~$s_0$ in the pattern space $P$, a new pattern~$s$ in the pattern space~$P$ is eventually created - first due to the movement of single cells and then due to the subsequent movement of single cells and agglomerates to new positions for a prescribed amount of time steps, see Algorithm~\ref{algo:each_iter}. The attractivity of potential new spots within the VNN is evaluated in each time step for the actual movement of the agglomerates and single cells.
% t is the number of neighbors a particle will have after moving to the potential new spot.
Maximizing attractivity involves trying all possible moves and selecting the one with the highest attractivity values.

First, all agglomerates within the domain $N^d$ are identified. These are \circlearound{B} and \circlearound{C} in the two-dimensional example as illustrated in Figure \ref{FIG:move}. The agglomerates are randomly ordered, and their potential movement within VNN(ag) is evaluated successively. Since the CA should compactify phase \circlearound{1}, each agglomerate's jumping is chosen to maximize the number of its direct neighbors. If several equally attractive new spots exist for an agglomerate, one of them is randomly selected.
 
Let us assume that \circlearound{B} is the first agglomerate to move in the two-dimensional example of Figure \ref{FIG:move}, and that it may move in a VNN of $1$ for the parameter choice $\sigma=2$. This means the agglomerate can remain in its actual position, moving to the left, right, upwards, or downwards. The most attractive new spot can here be achieved by moving downwards. After \circlearound{B} has moved, \circlearound{C} may move, but the number of neighbors will decrease if \circlearound{C} changes its position. Thus, it does not move to a new spot. 

After all the agglomerates have moved, all single cells (part of agglomerates or not) may move within their VNN to find new and more attractive spots. Again, the order of the movement of the single cells is random, and the single cells move such that they end up with a maximum amount of direct neighbors. If there are two equally beneficial moves, one of those is randomly selected. In the two-dimensional example illustrated in Figure \ref{FIG:move}, these movements are indicated in the middle picture by arrows for the parameter choice $\sigma=2$. 

Moving agglomerates and single cells is repeated several times (CA steps). In this sense, the forward model maps an initially disordered state vector $s_0 \sim U^{\theta}\{0,1\}$ into a pattern $s = F_\textup{CA}^{n^{*}}(\sigma)s_0$. This process naturally depends on the jumping parameter $\sigma \in \Pi$ and porosity parameter $\theta$.  We will assume that $\theta$ is fixed and use the parameter estimation method to recover parameter $\sigma \in \Pi$ from patterns. This is outlined below in Section~\ref{SEC:Parameter_estimation}. Note that due to the random origin of $s_0$, distinct patterns are emerging even for a given maximal size of the VNN (prescribed by the value of $\sigma$), and the inverse problem becomes a non-trivial task.

\subsection{Application of cellular automaton method for different parameter sets}
\begin{figure}\centering
 \begin{tabular}{ccc@{\;}c@{\;}c@{\;}c}
  & &\multicolumn{4}{c}{Porosity $\theta$} \\
  & & 0.3 & 0.5 & 0.7 & 0.9 \\
  \multirow{13}{*}{\rotatebox[origin=c]{90}{Jump parameter $\sigma$}} & \rotatebox[origin=l]{90}{1} & \includegraphics[width=0.2\textwidth,draft=false]{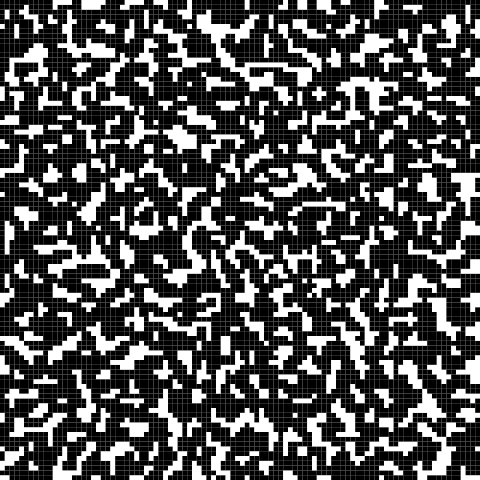} & \includegraphics[width=0.2\textwidth,draft=false]{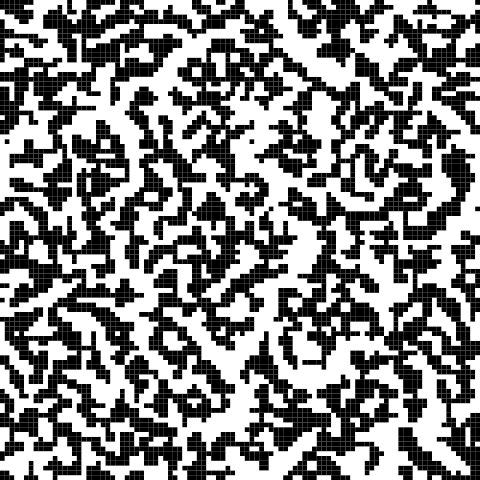} & \includegraphics[width=0.2\textwidth,draft=false]{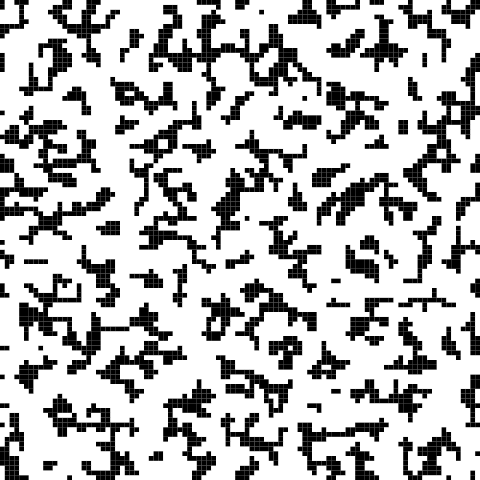} & \includegraphics[width=0.2\textwidth,draft=false]{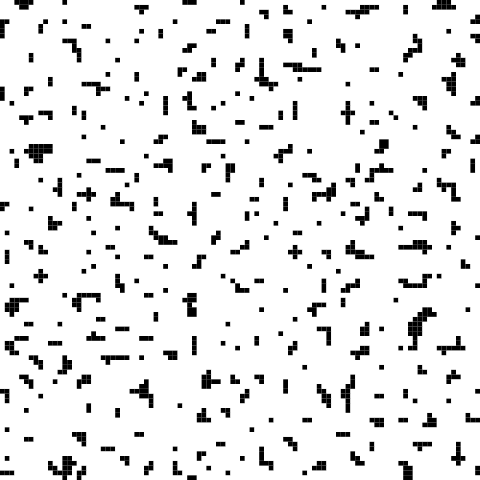} \\
  & \rotatebox[origin=l]{90}{5} & \includegraphics[width=0.2\textwidth,draft=false]{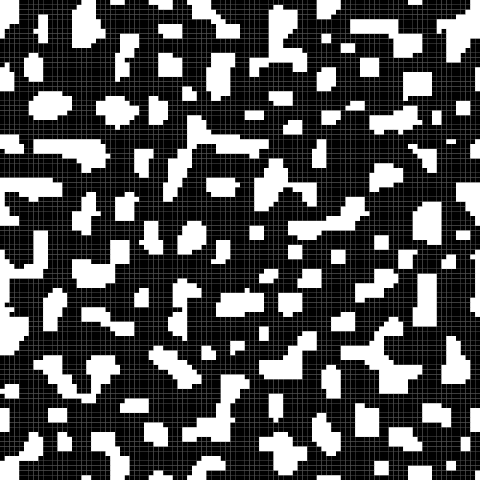} & \includegraphics[width=0.2\textwidth,draft=false]{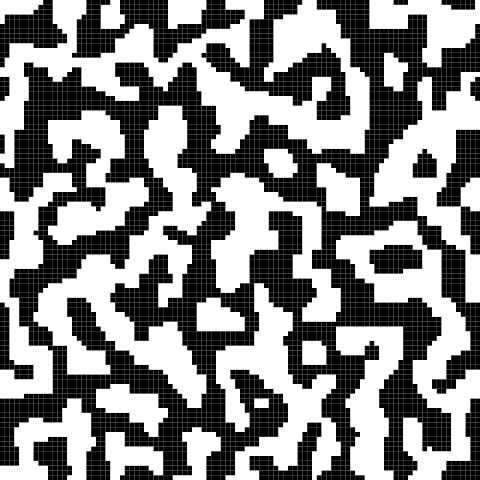} & \includegraphics[width=0.2\textwidth,draft=false]{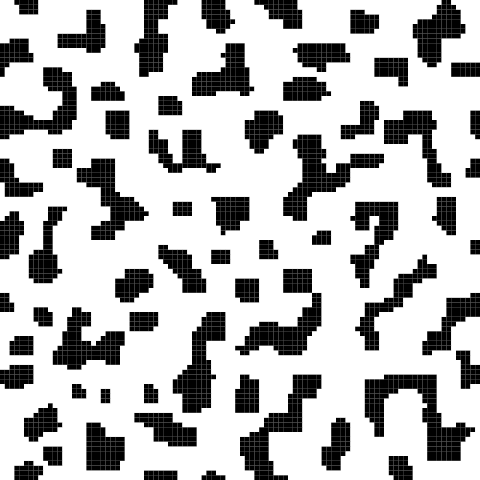} & \includegraphics[width=0.2\textwidth,draft=false]{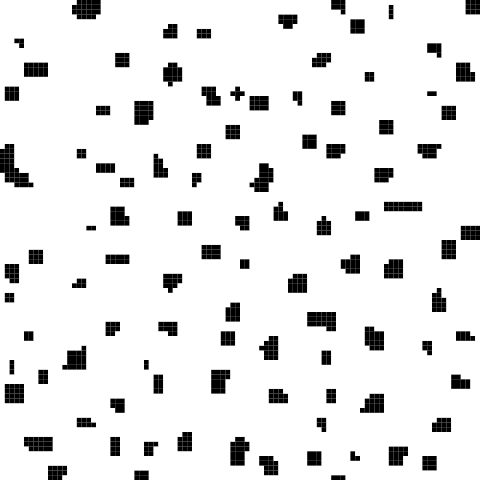} \\
  & \rotatebox[origin=l]{90}{10} & \includegraphics[width=0.2\textwidth,draft=false]{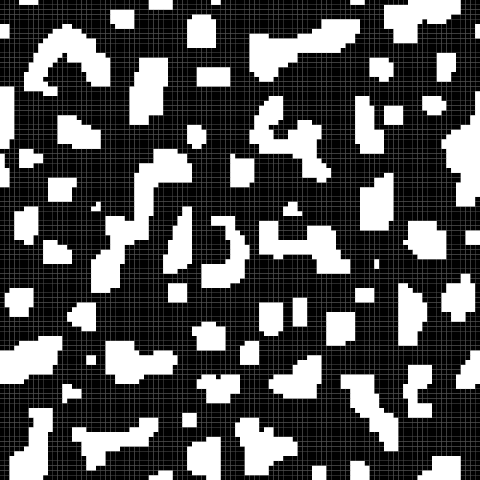} & \includegraphics[width=0.2\textwidth,draft=false]{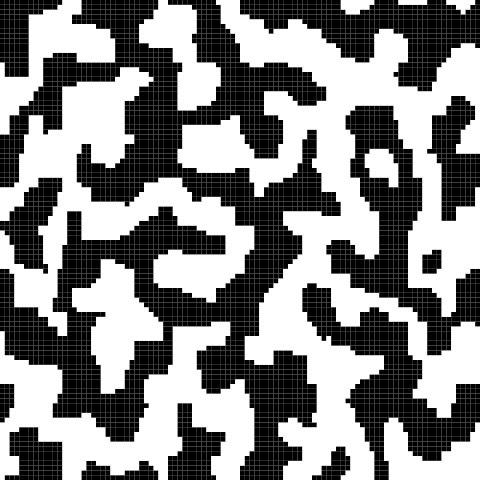} & \includegraphics[width=0.2\textwidth,draft=false]{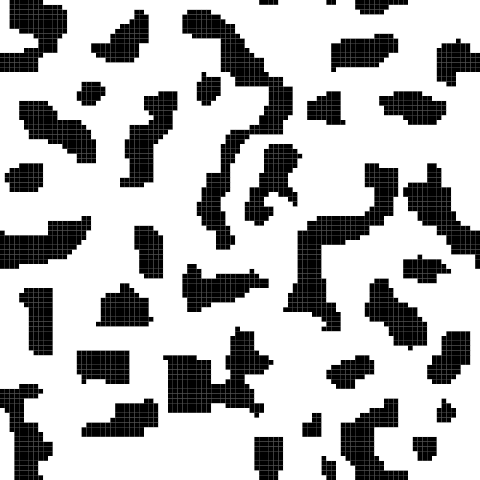} & \includegraphics[width=0.2\textwidth,draft=false]{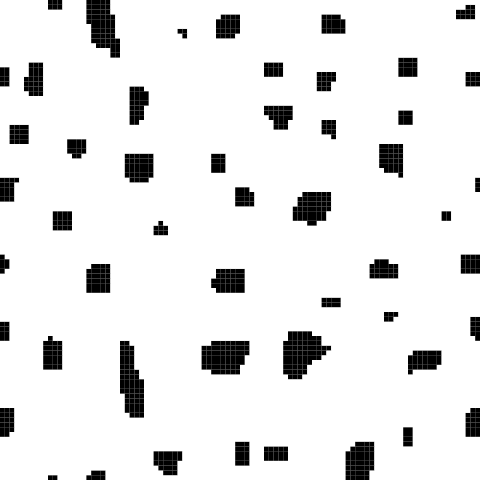} \\
  & \rotatebox[origin=l]{90}{15} & \includegraphics[width=0.2\textwidth,draft=false]{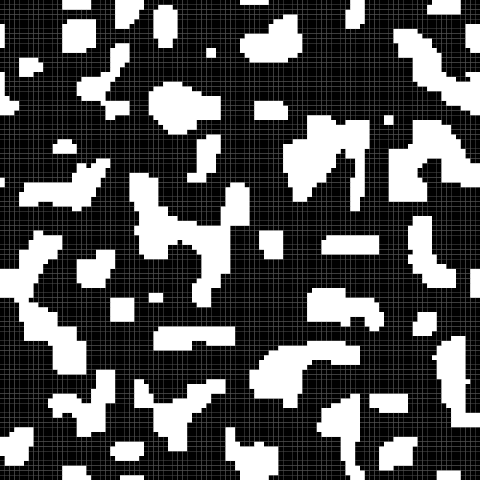} & \includegraphics[width=0.2\textwidth,draft=false]{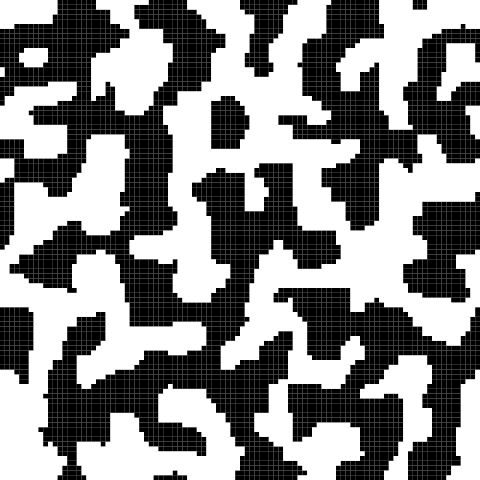} & \includegraphics[width=0.2\textwidth,draft=false]{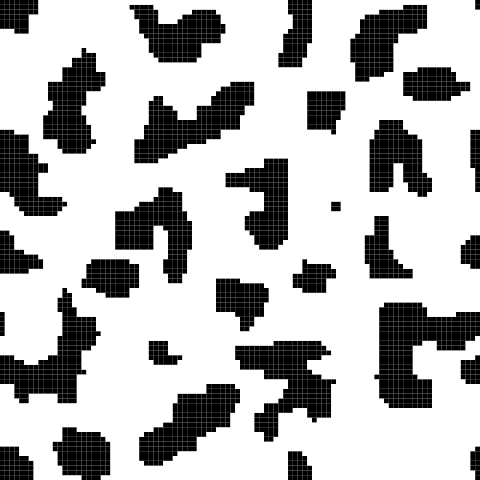} & \includegraphics[width=0.2\textwidth,draft=false]{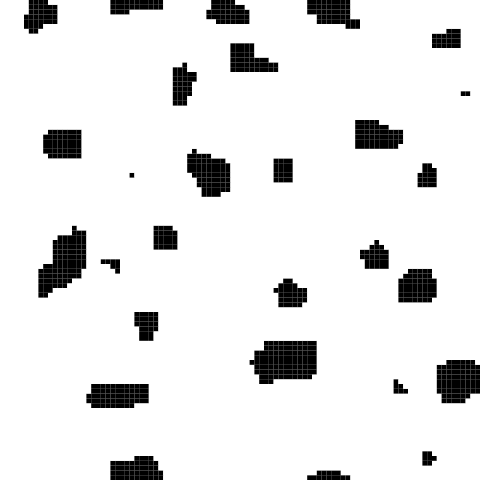}
 \end{tabular}
 \caption{Illustration of the domain consisting of $100 \times 100$ cells after five steps of the CA for varying porosity $\theta$ and jump parameter $\sigma$.}\label{fig:domain_porosity_jumpParameter}
\end{figure}

\begin{figure}\centering
 \begin{tabular}{ccc@{\;}cc@{\;}c}
  & &\multicolumn{2}{c}{$\theta = 0.5$} & \multicolumn{2}{c}{$\theta$= 0.9} \\
  & & $\sigma = 1$ & $\sigma = 5$ & $\sigma = 1$ & $\sigma = 5$ \\
  \multirow{23}{*}{\rotatebox[origin=l]{90}{Time step}} & \rotatebox[origin=c]{90}{0} & \includegraphics[width=0.2\textwidth,draft=false]{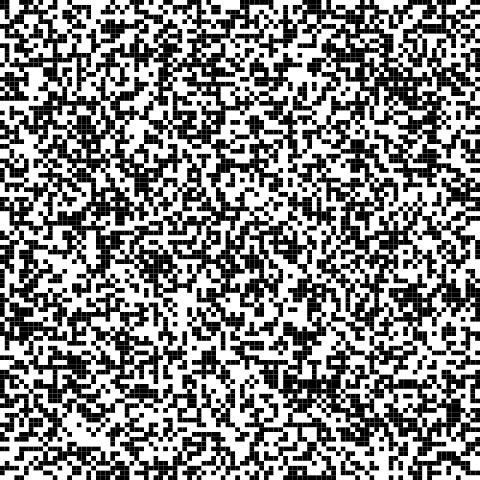} & \includegraphics[width=0.2\textwidth,draft=false]{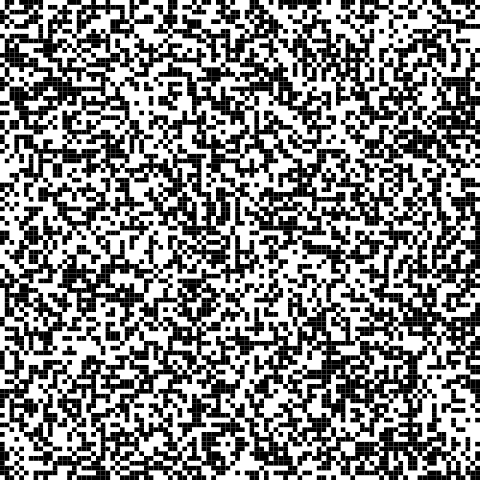} & \includegraphics[width=0.2\textwidth,draft=false]{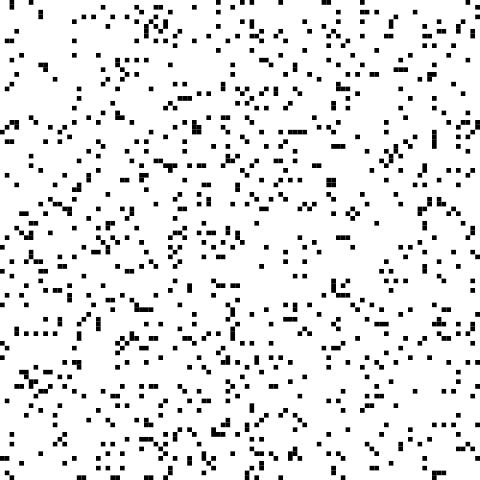} & \includegraphics[width=0.2\textwidth,draft=false]{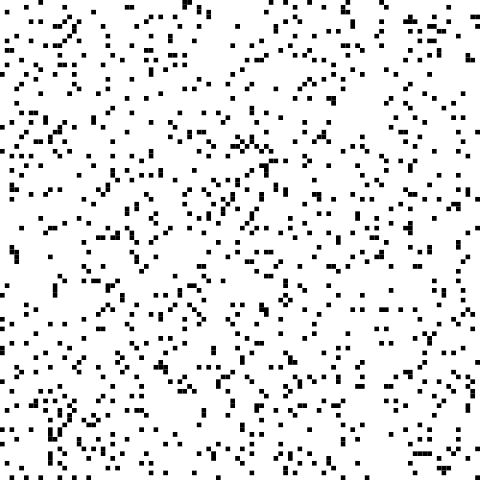} \\
  & \rotatebox[origin=l]{90}{1} & \includegraphics[width=0.2\textwidth,draft=false]{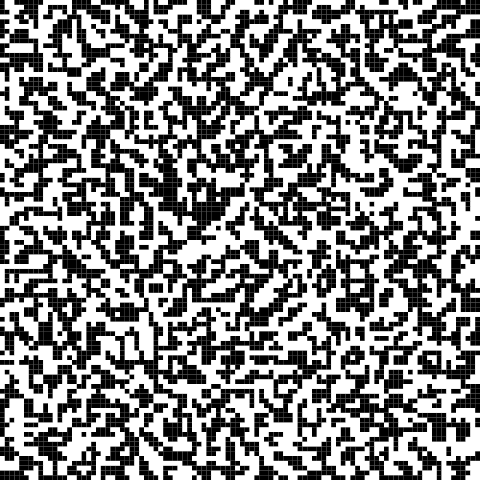} & \includegraphics[width=0.2\textwidth,draft=false]{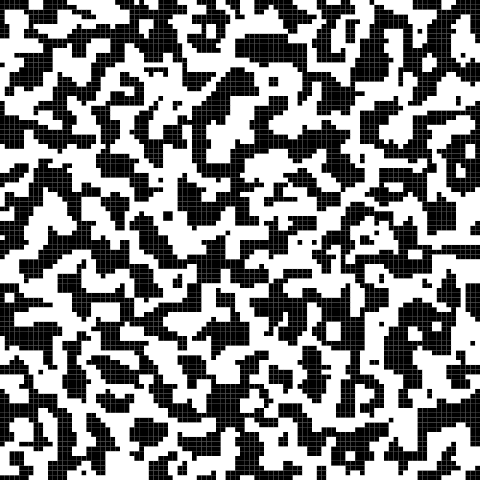} & \includegraphics[width=0.2\textwidth,draft=false]{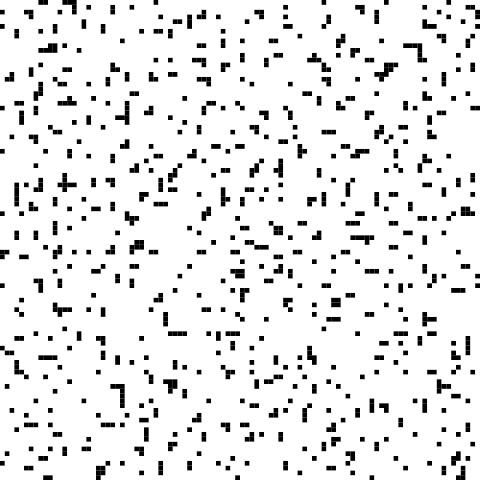} & \includegraphics[width=0.2\textwidth,draft=false]{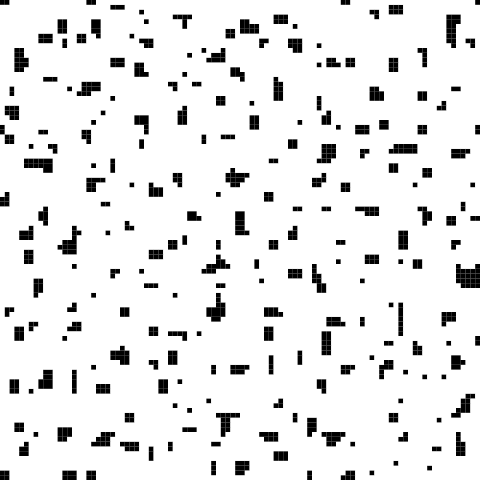} \\
  & \rotatebox[origin=l]{90}{5} & \includegraphics[width=0.2\textwidth,draft=false]{pictures/cam_illustration_python/cam_5_0.5_1.png} & \includegraphics[width=0.2\textwidth,draft=false]{pictures/cam_illustration_python/cam_5_0.5_5.png} & \includegraphics[width=0.2\textwidth,draft=false]{pictures/cam_illustration_python/cam_5_0.9_1.png} & \includegraphics[width=0.2\textwidth,draft=false]{pictures/cam_illustration_python/cam_5_0.9_5.png} \\
  & \rotatebox[origin=l]{90}{10} & \includegraphics[width=0.2\textwidth,draft=false]{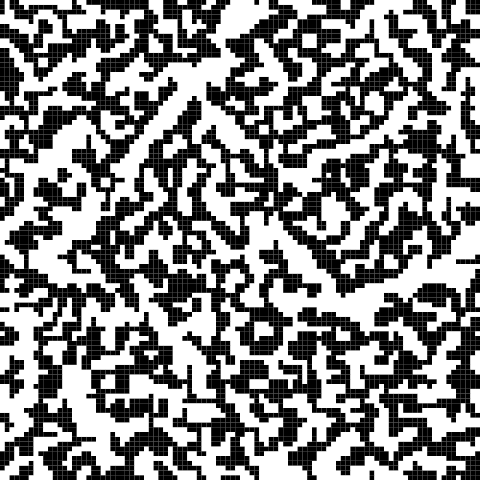} & \includegraphics[width=0.2\textwidth,draft=false]{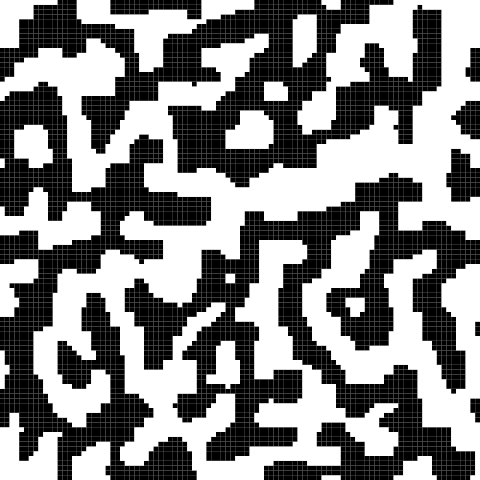} & \includegraphics[width=0.2\textwidth,draft=false]{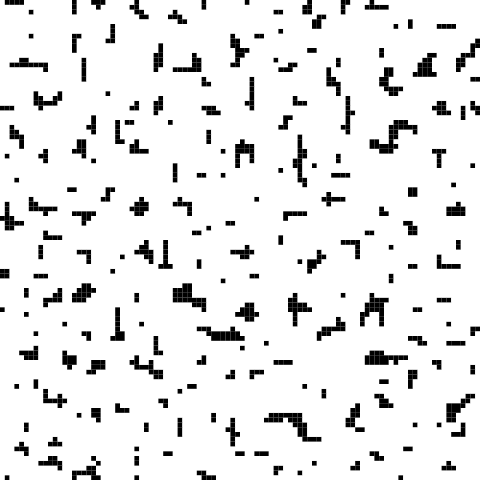} & \includegraphics[width=0.2\textwidth,draft=false]{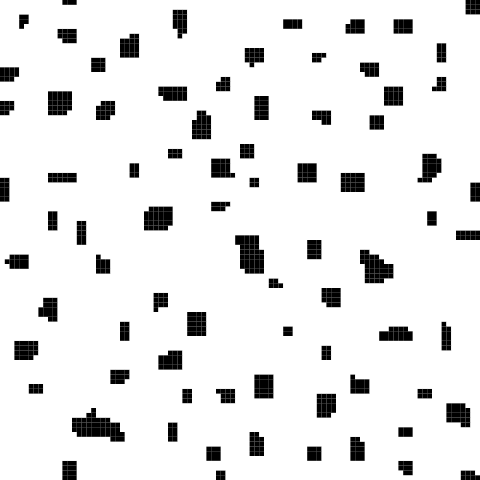} \\
  & \rotatebox[origin=l]{90}{100} & \includegraphics[width=0.2\textwidth,draft=false]{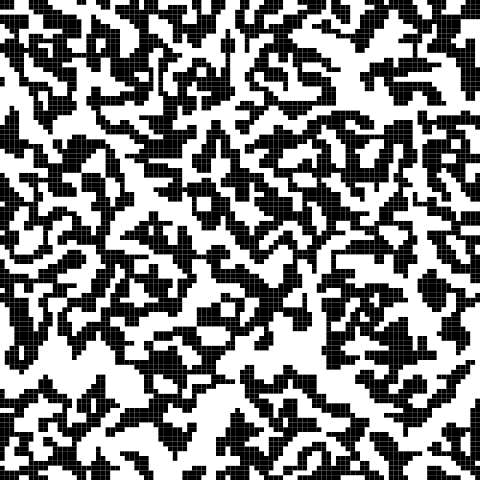} & \includegraphics[width=0.2\textwidth,draft=false]{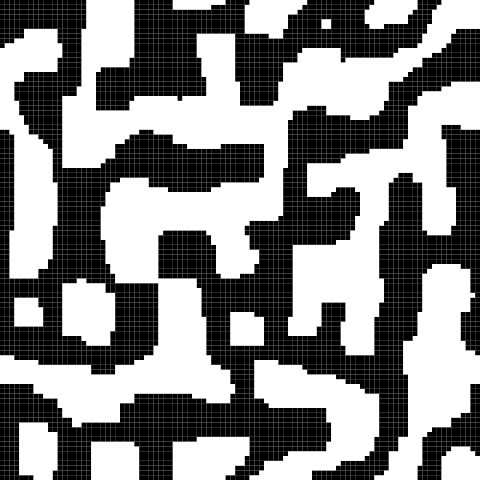} & \includegraphics[width=0.2\textwidth,draft=false]{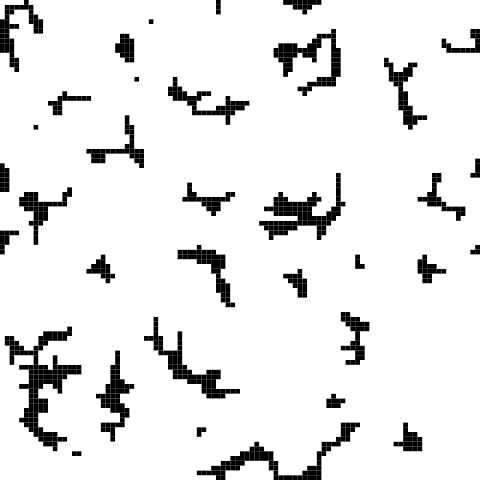} & \includegraphics[width=0.2\textwidth,draft=false]{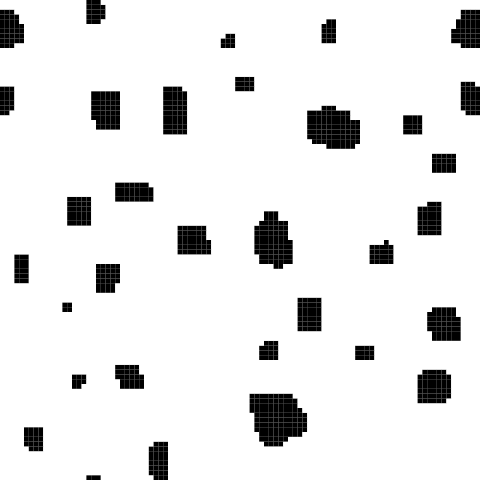} \\
  & \rotatebox[origin=l]{90}{1000} & \includegraphics[width=0.2\textwidth,draft=false]{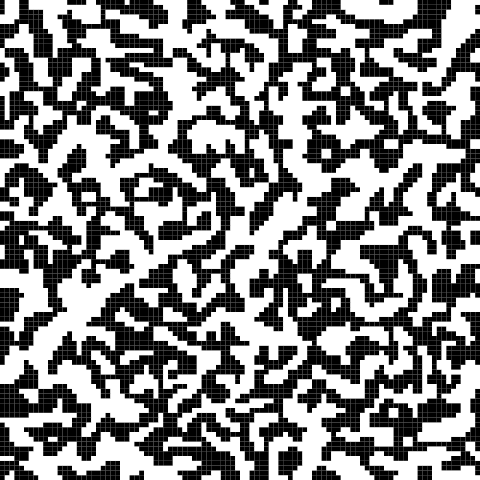} & \includegraphics[width=0.2\textwidth,draft=false]{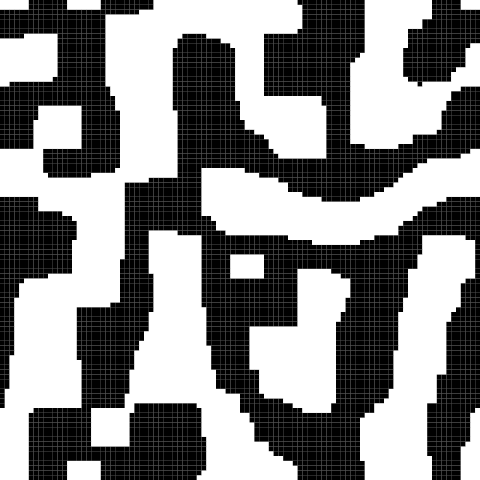} & \includegraphics[width=0.2\textwidth,draft=false]{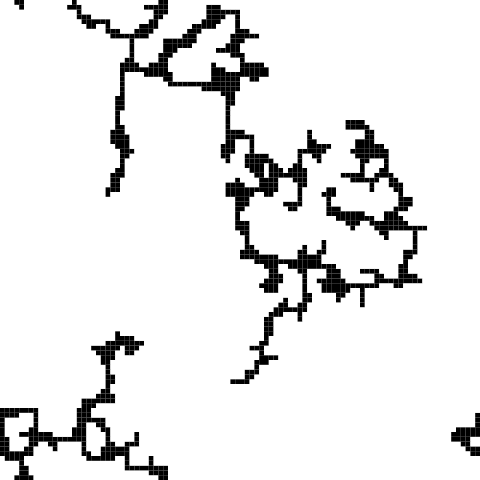} & \includegraphics[width=0.2\textwidth,draft=false]{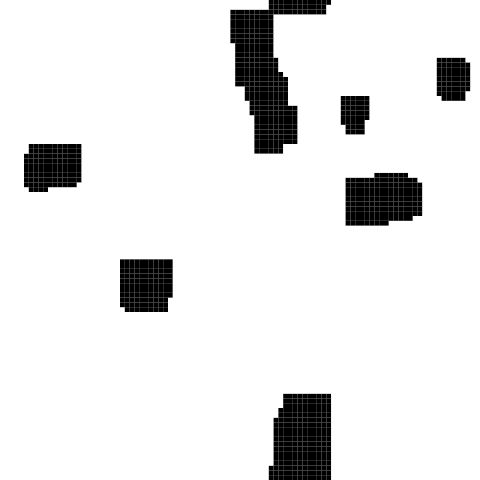}
 \end{tabular}
 \caption{Illustration of the time evolution of the domain consisting of $100 \times 100$ cells after steps $0, 1, 5, 10, 100$ and $1000$ of the CA for porosity $\theta \in \{0.5,0.9\}$ and jump parameter $\sigma \in \{1,5\}$.}\label{fig:time_porosity_jumpParameter}
\end{figure}

We apply the cellular automaton as introduced in Section~\ref{SEC:Jumping_rules} to illustrate the corresponding pattern formation for different choices of parameters. Starting from dispersed, randomly created structures, according to the CA jumping rules, single cells and agglomerates attract each other and finally form larger clusters and potentially connected structures. This process is illustrated in Figure~\ref{fig:domain_porosity_jumpParameter} for different choices of the jump parameter
\begin{equation*}
 \sigma \in \{1,5,10,15\},
\end{equation*}
and different porosities $\theta\in\{0.3,0.5,0.7,0.9\}$. The dynamic structure development concerning time is shown in Figure~\ref{fig:time_porosity_jumpParameter} for two different porosities $\theta \in \{0.5,0.9\}$ and jump parameters $\sigma \in \{1,5\}$. Distinct patterns emerge depending on the specific parameter choice. Larger porosities lead to dispersed structures, while larger jump parameters lead to blocky patterns. On the other hand, smaller porosities and smaller jump parameters induce card-house-type structures.

\begin{figure}\centering
\begin{tabular}{cccc}
    Timestep: 0 & Timestep: 1 & Timestep: 2 & Timestep: 3 \\
    \includegraphics[width=0.2\textwidth,draft=false]{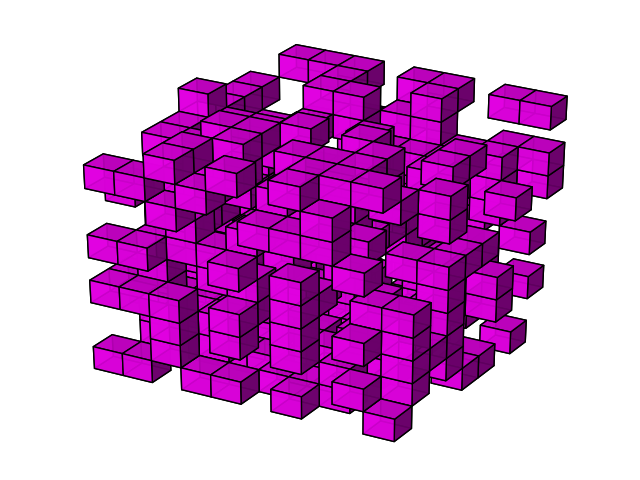} & \includegraphics[width=0.2\textwidth,draft=false]{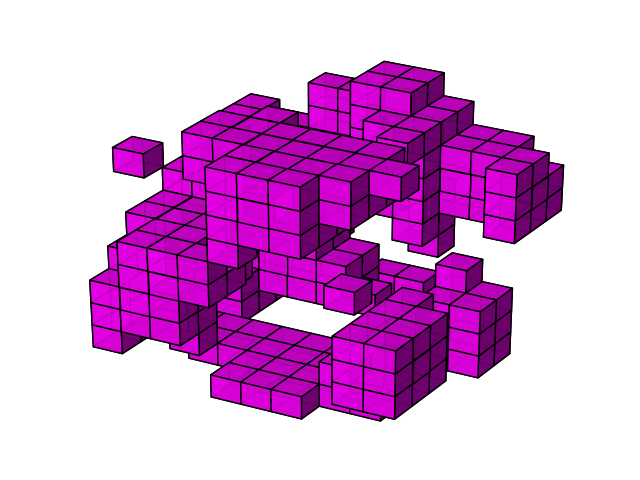} & \includegraphics[width=0.2\textwidth,draft=false]{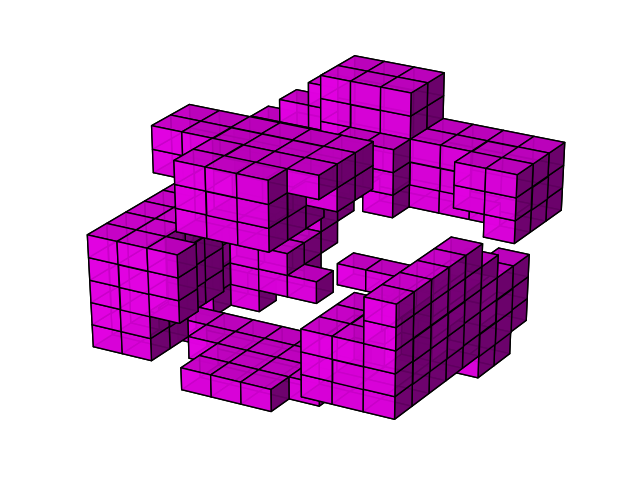} &
    \includegraphics[width=0.2\textwidth,draft=false]{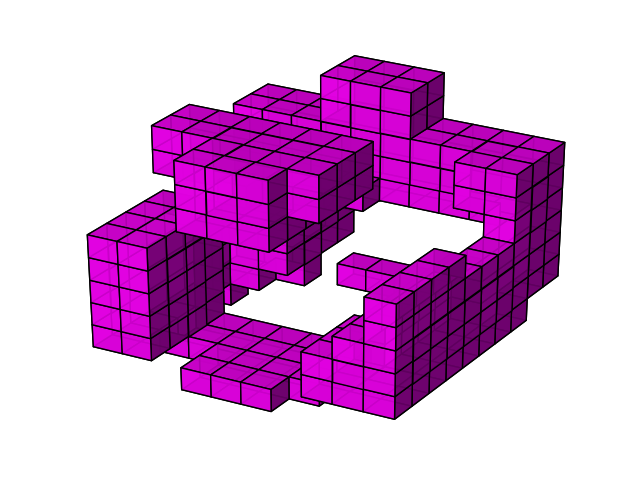}
\end{tabular}
\caption{Illustration of the time evolution of the domain consisting of $10 \times 10 \times 10$ cells after steps $0, 1, 2$ and $3$ of the CA for porosity $\theta = 0.7$ and jump parameter $\sigma = 5$.}
\label{FIG:CAM_3D}
\end{figure}

Besides the illustrations of the cellular automaton model for two spatial dimensions in Figure~\ref{fig:domain_porosity_jumpParameter} and~\ref{fig:time_porosity_jumpParameter}, the method can also be applied to model self-organization in three spatial dimensions as shown in Figure \ref{FIG:CAM_3D}. Likewise, it can also be applied in higher spatial dimensions using the implementation of \cite{RuppZL22}.
The resulting patterns are input for further analysis using parameter estimation methods outlined below in Section \ref{SEC:Parameter_estimation}.

\section{Parameter estimation method}\label{SEC:Parameter_estimation}
\begin{figure}\centering
 \includegraphics[width=0.99\textwidth,draft=false]{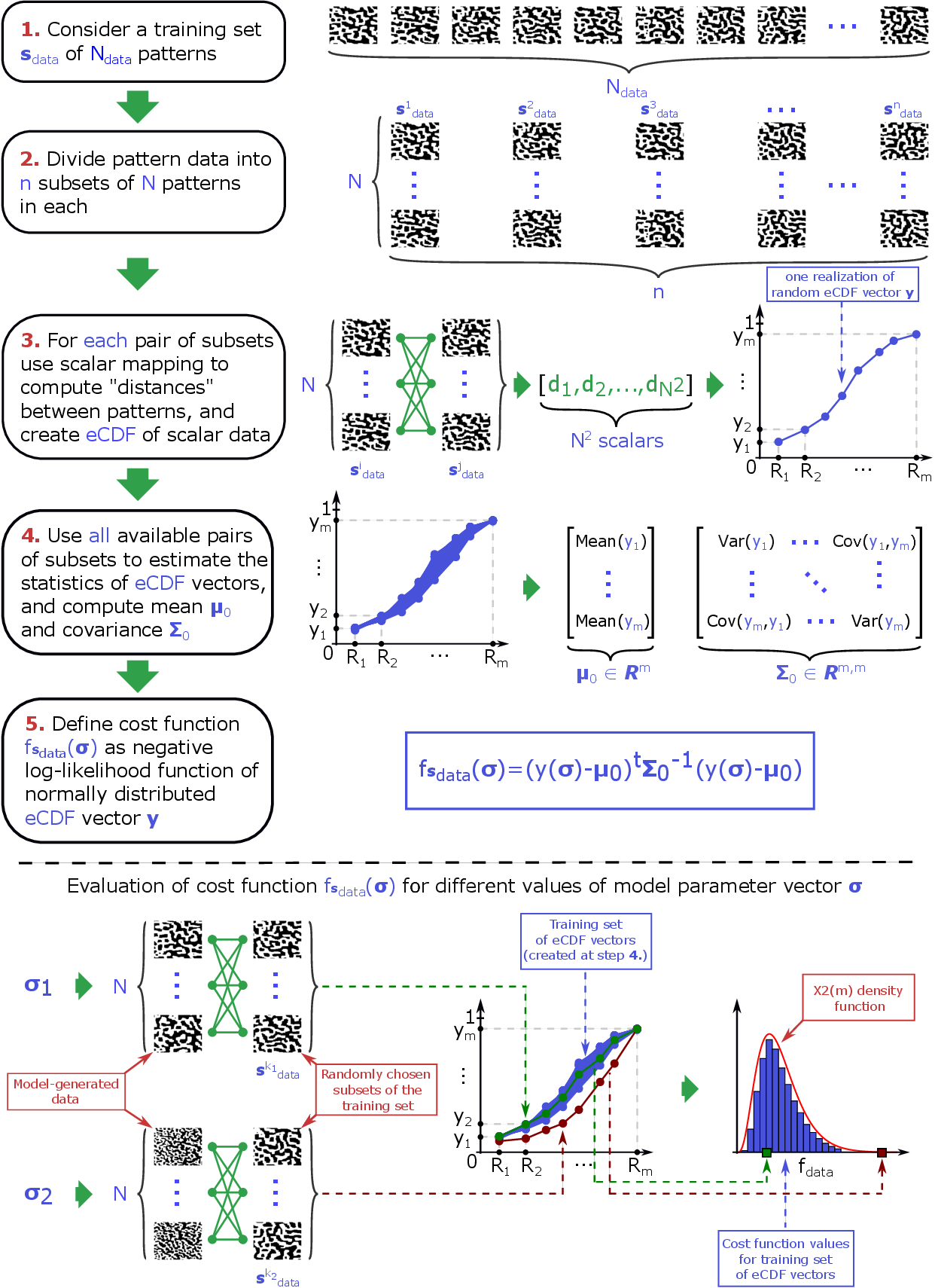} 
 \caption{Construction and evaluation of the cost function for parameter identification by pattern data.}\label{FIG:Workflow}
\end{figure}

We now introduce the parameter estimation method we apply to the results generated by the CA as outlined in Section~\ref{SEC:CAM}. Our method for parameter identification allows us to map a training set of patterns to a Gaussian distribution. This will enable us to define a statistical likelihood and use it as a cost function during the parameter identification. Our approach relies on emerged pattern data only, without using the information about the initial data.

\subsection{Background}
According to the central limit theorem,  the average of random variables with finite expected value and variance converges to the normal distribution. This allows for using a Gaussian likelihood for parameter estimation if enough repeated measurements are available. However, the mean may be rather uninformative, as quite different distributions can have the same mean (and higher moments).  The cumulative distribution function (CDF) can create more accurate statistics. In probability theory, Donsker's theorem is a functional extension of the central limit theorem. In this work, we build on generalizations of the Donsker theorem \cite{donsker1952}, which states that the cumulative distribution function of independent and identically distributed (i.i.d) scalar samples converges towards a Gaussian vector.
\begin{theorem}[Donsker, Skorokhod, Kolmogorov]
Let $F_n$ be the empirical distribution function of the sequence of i.i.d. random variables $ X_{1},X_{2},\ldots,X_{n}$ with distribution function F. Define the centered and scaled version of $F_n$ by
$$ G_{n}(x)={\sqrt {n}}(F_{n}(x)-F(x)).$$
The sequence of $G_n(x)$  converges in distribution to a Gaussian process $G$ with zero mean and the covariance is given by
$$ \operatorname{cov} [G(s),G(t)]=E[G(s)G(t)]=\min\{F(s),F(t)\}-F(s) F(t).$$
\end{theorem}
We use the theorem in an approximative form for finite data.  For i.i.d.\ scalar data with sample size $N$, the empirical distribution function (eCDF)  computed at  selected bin values $x_i$, $i=1,2,\dots,M$, becomes a $M$-dimensional Gaussian vector, with mean $\bm{F}_0 \in \mathbb{R}^M$ and covariance given by 
\begin{equation*}
 (\bm \Sigma_0)_{ij} = ({\rm min}((\bm{F}_0)_i,(\bm{F}_0)_j)- (\bm{F}_0)_i (\bm{F}_0)_j)/N, \quad i,j=1,\ldots,M.
\end{equation*}
  
The basic form of the Donsker theorem applies to i.i.d scalar situations. In our application, the data is not i.i.d. The covariance formula cannot be used then, but data or simulated eCDF vectors can estimate the covariance matrix. The Gaussianity still holds, assuming that conditions on weakly dependent data hold \cite{borovkova2001,neumeyer2004}.

Also, our data is inherently high dimensional, so a scalar-valued mapping must first be used to construct eCDF vectors; see \cite{KazarnikovH20,springer2019,haario2015} for earlier examples. We discuss the construction of such scalar-valued mappings below. 

Note the approximative character of the approach in a finite setting. As eCDF vectors are strictly limited in the interval $(0,1)$, the normality cannot hold close to the tails. In numerical applications, standard scalar normality tests can be used to verify the normality at the bin values used, and the $M$-dimensional  $\chi^2$ test can be used for the  Gaussianity of the eCDF vectors.

\subsection{Construction of the approach}
Let us represent the CA model as an abstract pattern formation model depending on a $p$-dimensional vector of model parameters $\bm{\sigma} \in \Pi\subset\mathbb{R}^p$. In our specific case, $\bm \sigma = \sigma \in \mathbb{N}$, thus $p = 1$, but the construction of the approach remains the same for a larger number of parameters. The output of each \enquote{forward model} run is a pattern $s(\bm{\sigma})\in P$, which corresponds to the final state after a prescribed number ${n^{*}}$ of applications of the cellular automaton forward operator $F_\textup{CA}(\sigma)$ as introduced in Section~\ref{SEC:CAM}: $s(\bm{\sigma}) = [F_\textup{CA}(\sigma)]^{n^{*}}s_0$, $s_0 \sim U^{\theta}\{0,1\}$. Applying the forward model for various choices of initial random state $s_0$ and fixed parameter(s) $\bm \sigma$, a set of patterns $\bm{s}(\bm{\sigma})$ is obtained. These patterns naturally change due to the variation of the model parameter(s) but additionally change even for the fixed model parameter(s) due to the random distribution of the two phases at an initial time and the randomness included in the CA steps. We aim to distinguish this internal variability from the systematic changes due to varying model parameters.  

More precisely, we want to find all the model parameters that fit a given training data, i.e., a set of patterns $\bm s_\textup{data}\subset \bm s(\bm \sigma)$, within the accuracy allowed by the data. To do so, we define a minimization problem in terms of a stochastic cost function
\begin{equation}\label{EQ:stoch_min_prob}
 f_{\bm s_\textup{data}} (\bm{\sigma}) \to \min \corr{\qquad \iff \qquad \bm{\bar\sigma} = \operatorname{arg min}_{{\bm\sigma}} f_{\bm s_\textup{data}} ( {\bm\sigma})},
\end{equation}
and consider any argument $\bm{\bar\sigma}$ that solves \eqref{EQ:stoch_min_prob} as a model parameter vector that corresponds to the training data set $\bm{s}_\textup{data}$. The remainder of this section is devoted to constructing $f_{\bm s_\textup{data}}$ step-by-step.

First, we specify what we mean by a solution to problem~\eqref{EQ:stoch_min_prob}. Due to the stochasticity of the model, a given model parameter corresponds to a distribution of solutions. We thus distinguish different model parameters by the respective distributions they produce. As the pattern data is high-dimensional, we define some measures to quantify the \enquote{distance} between two samples. For this purpose, we employ the training data to construct a statistical likelihood function that quantifies the variability within the data, i.e., gives a distribution of acceptable solutions. The basic idea is to define a \enquote{distance} mapping $\rho$, e.g., a scalar mapping, which compares two patterns $\bm{s}^+,\bm{s}^- \in P$ (possible choices of $\rho$ will be discussed below). The full statistics of $\rho$ are then used to produce a Gaussian likelihood based on/from the training data.

We next show how to construct the function $f_{\bm s_\textup{data}}$ for a given distance and all the training data pairs. To employ the training data statistically, we divide the set of patterns into $n$ subsets. We define a function that accepts two arguments: the sets of patterns $\bm s^+$ and $\bm s^-$ containing $N^+$ and $N^-$ patterns, respectively. Apart from the two arguments, it depends on two parameters: a radius $R > 0$ and the \enquote{distance} $\rho$:
\begin{equation}\label{eq:eCDF}
  C(\bm s^+,\bm s^-;R,\rho)=\frac{1}{N^+ \times N^-}\sum_{i=1}^{N^+} \sum_{j=1}^{N^-} \#(\rho(\bm{s}^+_{i},\bm{s}^-_{j})<R),
\end{equation}
where $\#$ denotes the discrete indicator function.

The radius $R > 0$ is used to create a vector $\bm y$ that encodes the similarity of two sets of patterns. Hence, we define a vector of bin values $(R_i)_{i=1}^m$, and set
\begin{equation*}
 \bm y(\bm s^+, \bm s^-) = \bm y(\bm s^+, \bm s^-; (R_i)_{i=1}^m, \rho) = \left( C(\bm s^+,\bm s^-;R_i,\rho) \right)_{i=1}^m.
\end{equation*}
This vector represents the eCDF of the set of values $\rho(\bm{s}^+_{i},\bm{s}^-_{j})$ evaluated at the bin values $(R_i)_{i=1}^m$.
 
We quantify the statistics (mean and variance) of $\bm y(\bm s^+, \bm s^-)$ among subsets $\bm s^+, \bm s^-$ of the whole training set $\bm s_\textup{data}$: For each subset pair, we receive $N^2$ scalar distance values, from which a single eCDF vector is computed. Repeating this for all distinct $n(n-1)/2$ pairs
\begin{equation*}
 \bm y^{k,l} =   \bm y(\bm s^k_\textup{data}, \bm s^l_\textup{data}) \in \mathbb R^m \qquad 0 \le k < l \le n,
\end{equation*}
we can evaluate the mean $\bm \mu_0 \in \mathbb R^m$ and the covariance $\bm \Sigma_0 \in \mathbb R^{m,m}$ of all  the  pairs of distinct eCDF vectors. 

As for most applications, the normality is numerically verified here as follows: We test for Gaussianity of the ensemble of vectors using the $\chi^2$-test (or scalar normality tests for the vector components at the bin values). For this purpose, we evaluate all the values of the negative log-likelihood function
\begin{equation*}
 (\bm y^{k,l} - \bm \mu_0)^T \bm \Sigma_0^{-1} (\bm y^{k,l} - \bm \mu_0),
\end{equation*}
and compare the resulting histogram against the density function of the distribution $\chi^2_M$ with $m$ degrees of freedom. 
  
To evaluate the cost function at a new parameter value $\bm{\sigma}$, we simulate the CA model $N$ times using $\bm{\sigma}$ and denote the collection of patterns by $\bm s(\bm \sigma)$. The eCDF vector
\begin{equation}
\bm y(\bm{\sigma}) = \bm y(\bm s(\bm\sigma), \bm s^k_\textup{data})
\label{eq:obs_vector}
\end{equation}
can then be computed for one randomly selected $k\in\{1,2,\dots,n\}$, and the likelihood value is evaluated as
\begin{equation}
 f_{\bm s_\textup{data}} (\bm{\sigma}) = (\bm y(\bm{\sigma}) - \bm \mu_0)^T \bm \Sigma_0^{-1} (\bm y(\bm{\sigma}) - \bm \mu_0).
 \label{eq:cil_likelihood}
\end{equation}

To summarize, the 'forward model' is given by the CA algorithm that produces the collection of patterns by $\bm s(\bm \sigma)$ for a given value of $\bm{\sigma}$.  The observation operator is defined by applying formula \eqref{eq:obs_vector} that maps the set of model-generated patterns $\bm{s}(\bm{\sigma})$ to an eCDF vector $\bm{y}(\bm{\sigma})$. The 'inverse problem' of parameter estimation is solved by evaluating the stochastic cost function $f_{\bm s_\textup{data}}$ for a given value of $\bm{\sigma}$ by using expression \eqref{eq:cil_likelihood} and finding the minimum of it concerning $\bm{\sigma}$. 

Note that in the examples of the present work, we only work with one parameter, the integer-valued $\sigma$, so the parameter estimation can be performed simply with a direct search. 

\subsection{Numerical implementation}

%\subsubsection{Observation operator}

\subsubsection{Choice of bins}\label{SEC:Bins}

The bin values for the eCDF vectors can be selected in various ways. Here, we use the following approach: We first use the first two subsets of the training data and determine the minimum and maximum of the function $C$ as defined in \eqref{eq:eCDF} (concerning $R$). Next, we uniformly split the respective interval into $50$ possible bin values. This allows us to represent the shape of the eCDF curve. However, in the numerical application, such dense coverage might result in an unwanted correlation between neighboring bin values. Therefore, from these preliminary bins, we choose a smaller subset of $n_{bin}$ values, which we select by an inverse CDF method: to get the bin values on the x-axis, a set of $n_{bin}$ linearly spaced values on the y-axis are mapped to the x-axis by the inverse CDF (quantile) function, using the mean of the preliminary CDF vectors.
Additionally, a cut-off parameter of $0.1$ is used to step away from the CDF function's range $[0,1]$ boundaries. This allows us to exclude bins with possibly prohibitively small variabilities, which might result in singularities in the covariance matrix of the underlying Gaussian distribution. After that, the Gaussianity of both bin configurations (the preliminary and the selected sparse one) can be evaluated.
\subsubsection{Choice of characteristics}\label{SEC:Distance}
The algorithm of \cite{KazarnikovH22} employs several norms to characterize the distance between two patterns in continuous-valued images. As the present setting is discrete, with binary-valued patterns, we use the $L^1$-norm, i.e., we define the distance function $\rho(\bm{s}^+,\bm{s}^-) = \| \bm{s}^+ - \bm{s}^-\|_{L^1}$ between two patterns $\bm{s}^+,\bm{s}^-$.

However, various other candidates for characteristics are reasonable for the discrete CA patterns. Here, we use the particle size distribution (PSD) and the average particle size 
(APS), which are frequently considered in soil science~\cite{RuppGMPRRT19}. For the former case, let us consider a vector containing the sizes of all agglomerates $\boldsymbol \mu(\textup{ag})$.

% The particle size distribution is the concatenated vector
% \begin{equation*}
%  (\underbrace{1,\dots,1}_{\textup{Number of single cells}},\boldsymbol \mu(\textup{ag})).
% \end{equation*}

Let us consider the first two patterns in the example of Figure~\ref{FIG:move}. The PSD for the left figure (pattern $s^+$) reads $(1,1,1,4,4)$ as three single cells and two aggregates of size four are present. In the middle figure (pattern $s^-$), two single cells and one agglomerate of size 9 are present, i.e., the PSD reads $(1,1,9)$.
The average particle size is the mean of the particle size distribution, i.e., it is defined as the arithmetic average of the sizes of all particles, i.e., single cells and agglomerates. In our example this leads to APS$(s^+)=2.2$ and APS$(s^-)=3.67$, respectively. The distance function is given by
\begin{equation*}
 \rho(\bm{s}^+,\bm{s}^-) = | \textup{APS}(\bm{s}^+) - \textup{APS}(\bm{s}^-) |.
\end{equation*}
As our approach is based on the statistical distribution of the CDF functions of scalar data, we can also employ the particle size distribution directly by forming its empirical CDF. Although the particle size distribution could be computed for each pattern separately, we compute the particle size distributions of all pairwise combinations of patterns.
In this case, the ``distance'' is the concatenated vector,
\begin{equation*}
 \rho(\bm{s}^+,\bm{s}^-) = \left( \textup{PSD}(\bm{s}^+), \textup{PSD}(\bm{s}^-) \right) 
\end{equation*}
and the indicator function in \eqref{eq:eCDF} must be replaced by a counting function. The concatenation of the pattern from our example leads to $(1,1,1,4,4,1,1,9)$. For the choice of $R=2$, the evaluation of $\rho$, for instance, leads to 5 counts.
This has the advantage of producing more eCDF vectors, stabilizing the numerical estimation of the mean and covariance of the likelihood. 

\subsubsection{Application of the eCDF method to the base setup}\label{SEC:results_base}
We perform numerical experiments to demonstrate how our parameter estimation method can be applied to identify parameters of the CA model from synthetic (model-generated) data. We first consider a basic experimental setup, which is defined as follows:
\begin{itemize}
 \item The CA is run on a two-dimensional domain of size $50 \times 50$ with porosity $\theta = 0.7$ and jump parameter $\sigma = 5$.
 \item The training set of patterns is obtained by running five iterations of the CA model with random initial data (4000 = 40$\times$100 realizations).
\end{itemize}

For this basic experimental setup, we create the $L^1$ likelihood and estimate the model parameters using the $L^1$ norm. 
 
\begin{figure}\centering
 \includegraphics[trim=125 20 125 30, clip, width=.99\textwidth, draft=false]{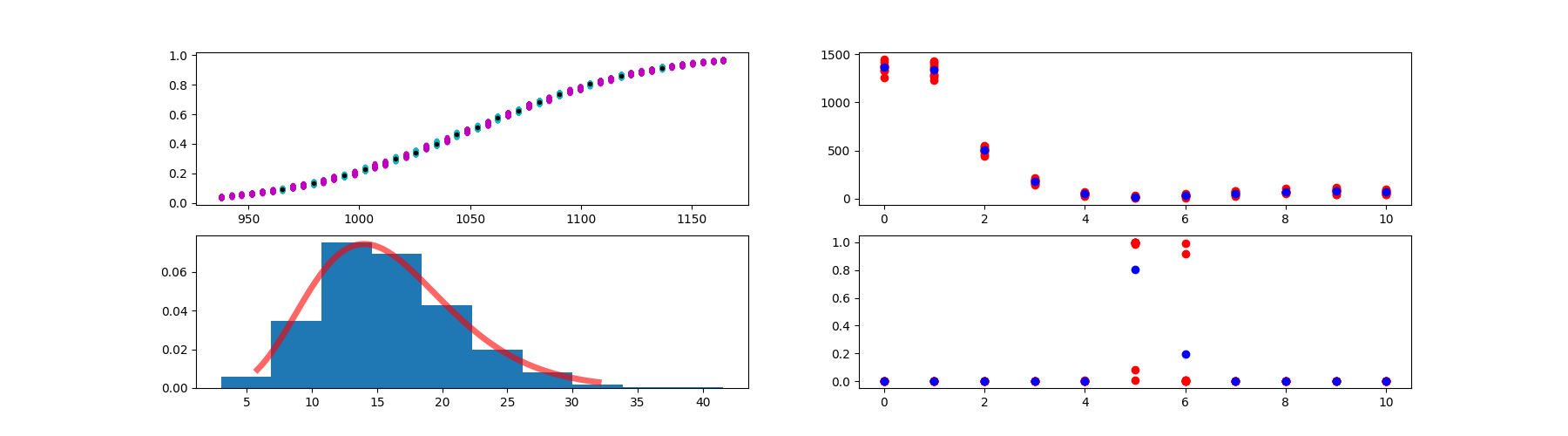}
 \caption{Construction of the statistical likelihood for basic experimental setup and its evaluation for different values of jump parameter $\sigma$. On the top left is the distribution of the eCDF curves for dense bin values (purple) and selected bins only (light blue).
 % Here, black dots denote the respective mean values of eCDF vectors over all available pairs.
 Bottom left: Gaussianity test by $\chi^2$ criterion for selected bin values.  Top right: repetitive evaluation of the negative log-likelihood (cost function) for integer values of jump parameter $\sigma$ on the interval $[0,10]$. Here, cost function values are plotted in red, while blue dots denote the average overall evaluations. Bottom right: normalized likelihood values (red) and averaged values over evaluations (blue).}\label{FIG:base_results}
\end{figure}

Following the procedure outlined in Section~\ref{SEC:Parameter_estimation} and illustrated in Figure~\ref{FIG:Workflow}, we create a statistical likelihood from the training data. We split the training set into 40 subsets with 100 samples in each. Next, for every pair, we compute distances between the respective patterns in terms of $L^1$-norm and compute the eCDF of the individual scalar data. Here, we successively select bins for the eCDF vectors using the algorithm described in Section~\ref{SEC:Bins}. The selection is illustrated in Figure~\ref{FIG:base_results} (top left). 

Next, we check the Gaussianity of the selected bins as outlined in Section~\ref{SEC:Parameter_estimation}, which is illustrated in the bottom left part of Figure~\ref{FIG:base_results}. We evaluate the negative log-likelihood for integer values of the jump parameter $\sigma$ on the interval $[0,10]$, as is shown in Figure~\ref{FIG:base_results} (top right). Here, the red dots represent the evaluation of this value 100 times, and the blue dots highlight the average values of the red dots. The minimum negative log-likelihood values are achieved at $\sigma=5$. Finally, the proper probabilistic interpretation is obtained by the normalized likelihood values in Figure~\ref{FIG:base_results} (bottom right)---with the same understanding of red and blue dots. Thus, our approach can properly distinguish patterns corresponding to different values of $\sigma$ for this setup.

\section{Results}\label{SEC:Results}
Next, we investigate the robustness of the parameter estimation method as introduced in Section~\ref{SEC:Parameter_estimation} concerning the number of bins, the domain size, and the number of time steps (iterations) in the CA model. Here, we change only one quantity at a time, while the other ones are fixed to the default values as prescribed by the basic experimental setup in the previous Section. Finally, we analyze the impact of including additional CA-specific characteristics to the scheme of the parameter estimation method, which is discussed in Section \ref{SEC:multi_feature}.
\subsection{Varying number of bins}
We first study the robustness of our parameter estimation method for the number of selected bins (dimension of eCDF vectors). We repeat the experiments described in Section~\ref{SEC:results_base} for varying bins between 6 and 26 while keeping the other parameters fixed. The results were statistically identical to the ones shown in Figure~\ref{FIG:base_results} for all considered cases. From this result, we conclude that the approach allows for relatively large flexibility concerning the number of bins once the numerical stability considerations discussed in Section~\ref{SEC:Bins} are considered.

\begin{figure}\centering
 \subfloat[$10 \times 10$]{
 \includegraphics[trim=125 20 125 30, clip, width=.98\textwidth,draft=false]{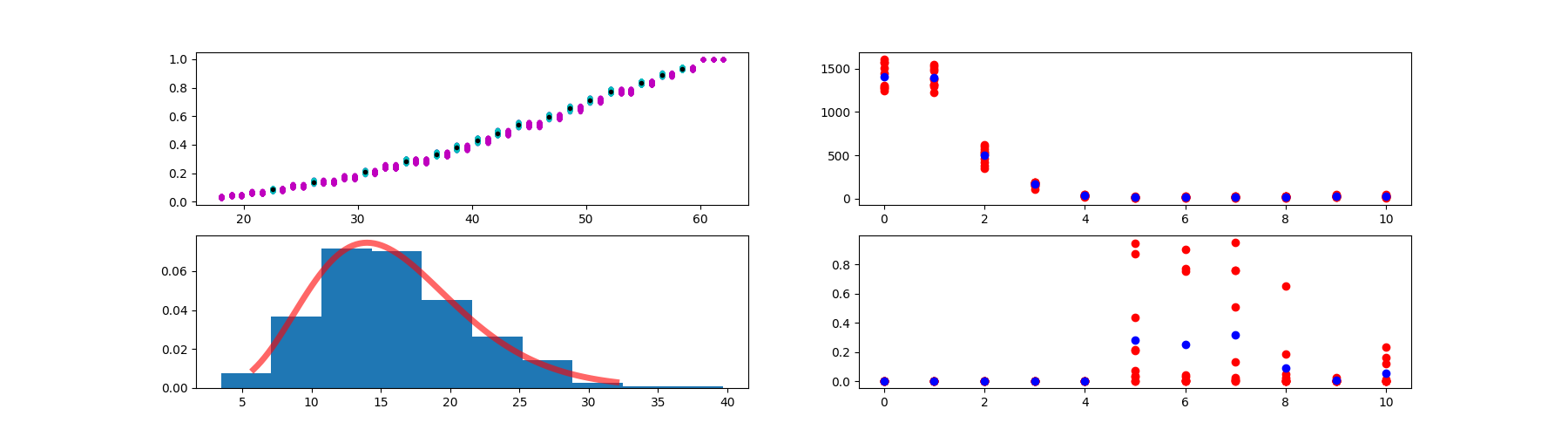}}\\
 \subfloat[$25 \times 25$]{
 \includegraphics[trim=125 20 125 30, clip, width=.98\textwidth,draft=false]{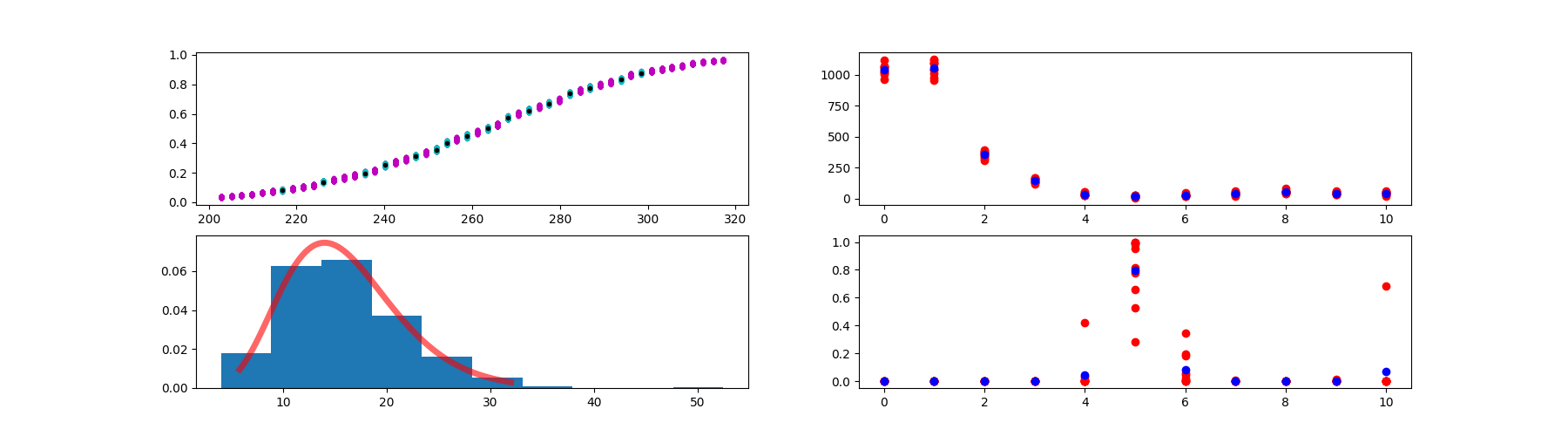}}\\
 \subfloat[$100 \times 100$]{
 \includegraphics[trim=125 20 125 30, clip, width=.98\textwidth,draft=false]{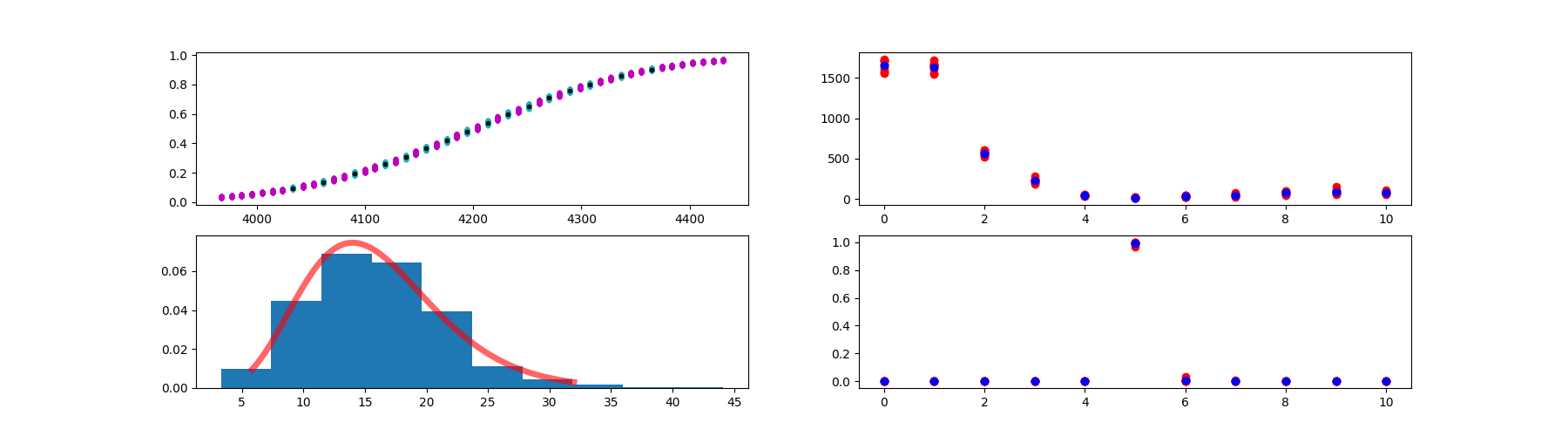}}\\
 \caption{ Construction of the statistical likelihood for various domain sizes and its evaluation for different values of jump parameter $\sigma$. The layout of sub-figures is identical to  Figure~\ref{FIG:base_results}.}
 \label{FIG:domain_sizes}
\end{figure}
\subsection{Varying domain sizes}
We now consider different two-dimensional domain sizes, i.e., we perform the procedure of Section \ref{SEC:results_base} for domains of size $N^2=10 \times 10$, $25 \times 25$, and $100 \times 100$. A simulation conducted on a small domain provides less information than a large one but keeps the porosity fixed. This is underpinned by the numerical experiments illustrated in Figure~\ref{FIG:domain_sizes}.

Considering first the $10 \times 10$ domain, our algorithm can correctly identify the true parameter $\sigma=5$. However, the minimum is not very pronounced, so the accuracy is relatively low. For domain sizes of $25 \times 25$,  $50 \times 50$ (basic experimental setup, see Figure \ref{FIG:base_results}), and $100 \times 100$, we observe smooth eCDF shapes again. Thus, the parameter estimation method works accurately and successfully. The increase in domain size leads to a more pronounced minimum at $\sigma=5$. Therefore, the algorithm is more confident in identifying the correct jump parameter if the domain size and available information increase.

\begin{figure}\centering
\subfloat[zero steps]{
 \includegraphics[trim=125 20 125 30, clip, width=.98\textwidth,draft=false]{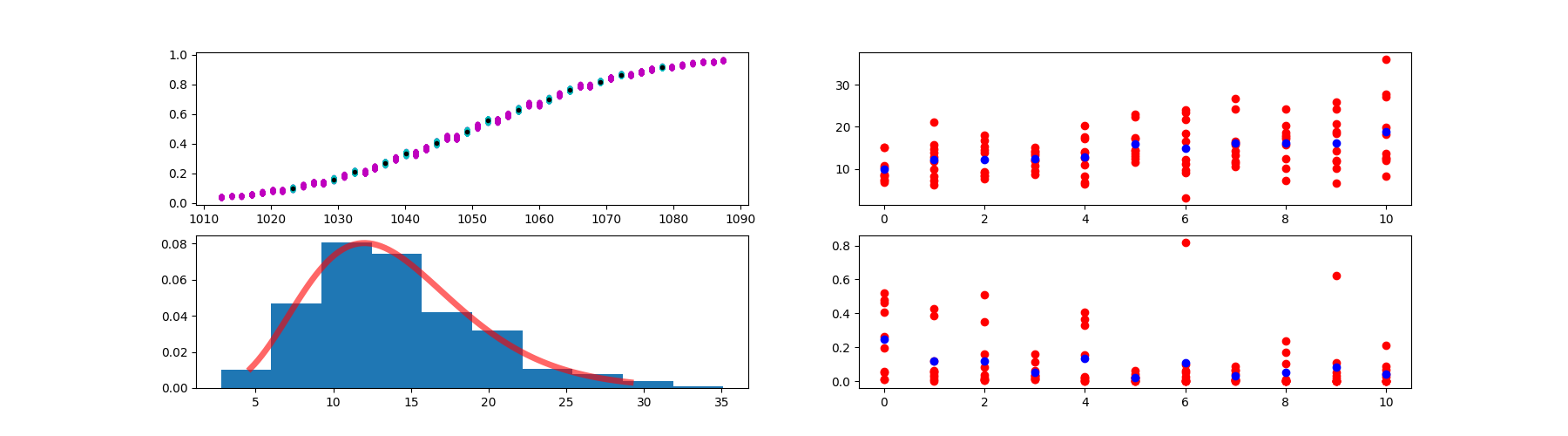}}\\
 \subfloat[$10$ steps]{
 \includegraphics[trim=125 20 125 30, clip, width=.98\textwidth,draft=false]{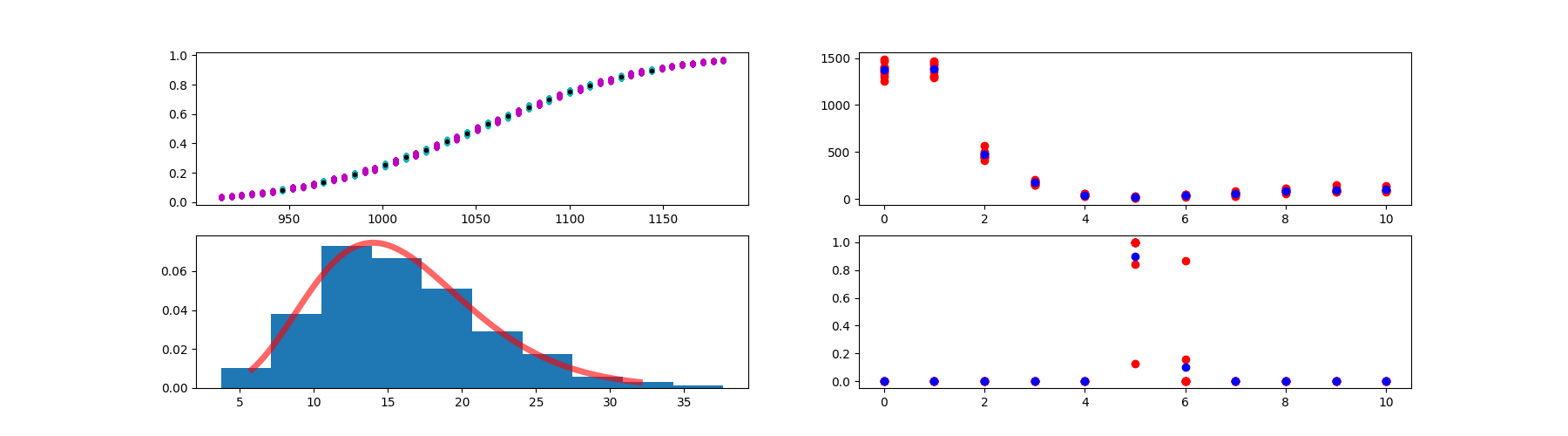}}\\
 \subfloat[$25$ steps]{
 \includegraphics[trim=125 20 125 30, clip, width=.98\textwidth,draft=false]{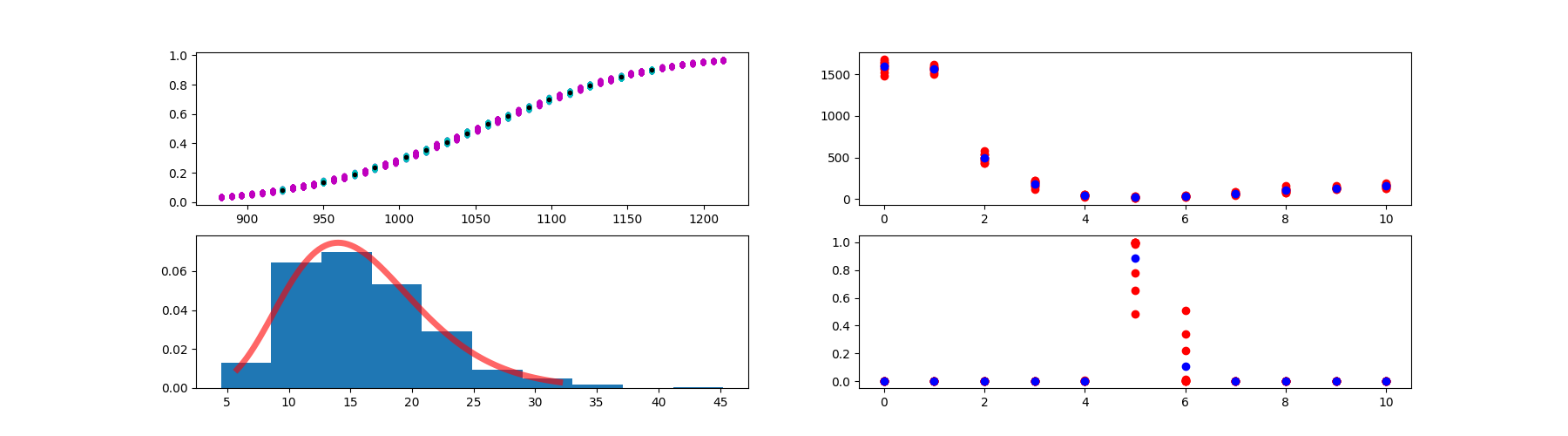}}\\
 \subfloat[$50$ steps]{
 \includegraphics[trim=125 20 125 30, clip, width=.98\textwidth,draft=false]{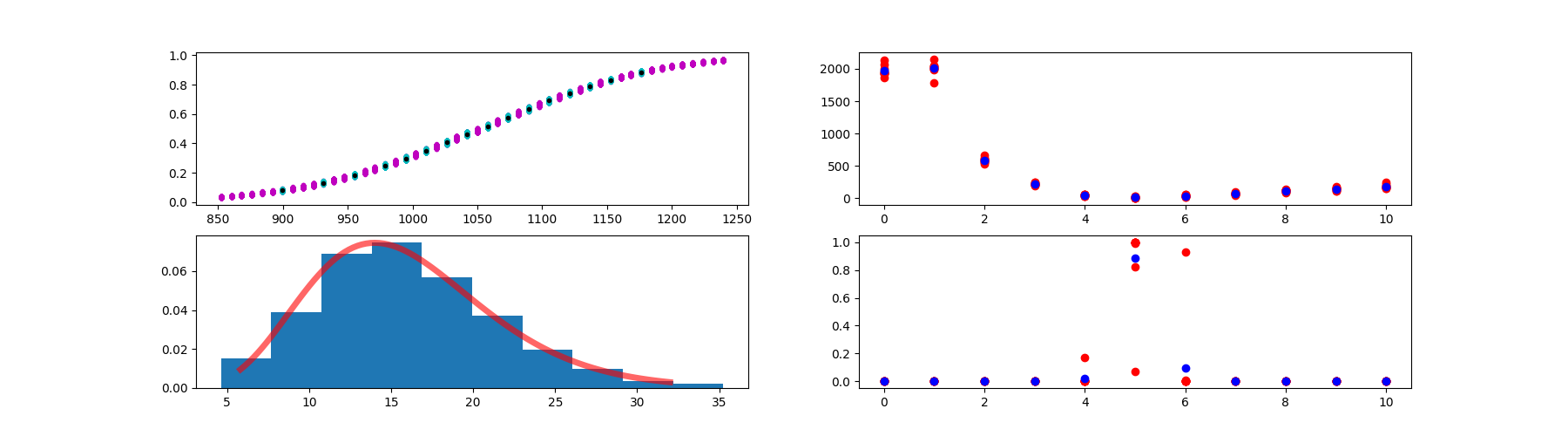}}
 \caption{Construction of the statistical likelihood for various CA iterations, and its evaluation for different values of jump parameter $\sigma$. The layout of sub-figures is identical to Figure~\ref{FIG:base_results}.}\label{FIG:time_steps}
\end{figure}
\subsection{Varying number of CA iterations}
The patterns generated by our CA model are not stationary since initially fragmented particles eventually attract each other and self-organize into larger, connected structures. Thus, depending on how many model iterations were used to create a pattern, it can be more or less clustered; see also illustration in Figure~\ref{fig:time_porosity_jumpParameter}. Here, we study how this factor affects our ability to perform parameter identification. The results are illustrated in Figure~\ref{FIG:time_steps}. We start by considering the trivial case of zero iterations. In this case, the dispersed initial state is independent of $\sigma$; thus, we can construct the likelihood without any problems. Still, we naturally cannot detect any difference between different values of $\sigma$. However, the parameter identification works without issues for all other considered cases: $1$, $5$ (base experimental setup, see Figure \ref{FIG:base_results}), $10$, $25$, or $50$ iterations. This indicates that the number of iterations does not significantly influence the quality of our parameter estimation method.

\begin{figure}\centering
\subfloat[$L^1$-distance]{
 \includegraphics[trim=125 20 125 30, clip, width=.7\textwidth,draft=false]{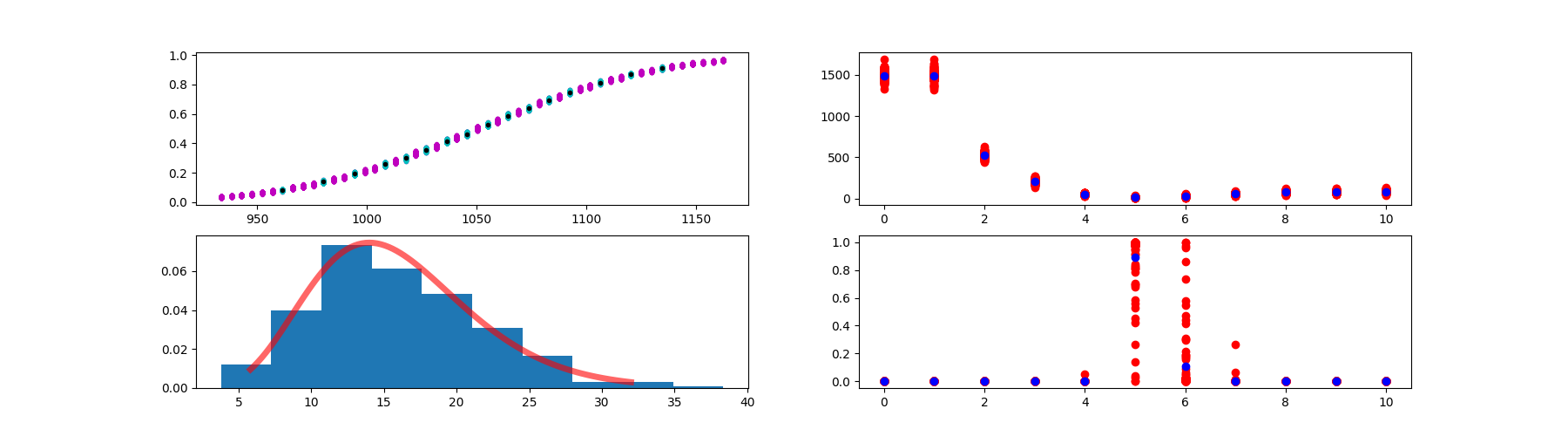}}\\
 \subfloat[absolute difference of average particle sizes]{
 \includegraphics[trim=125 20 125 30, clip, width=.7\textwidth,draft=false]{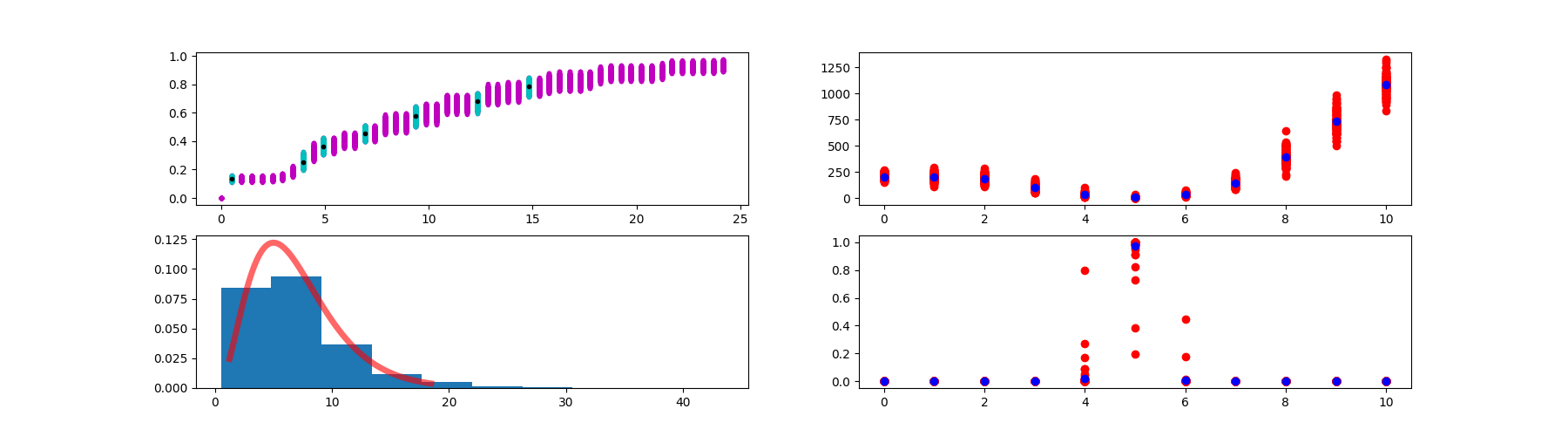}}\\
 \subfloat[$L^1$-distance + absolute difference of average particle sizes]{
 \includegraphics[trim=125 20 125 30, clip, width=.7\textwidth,draft=false]{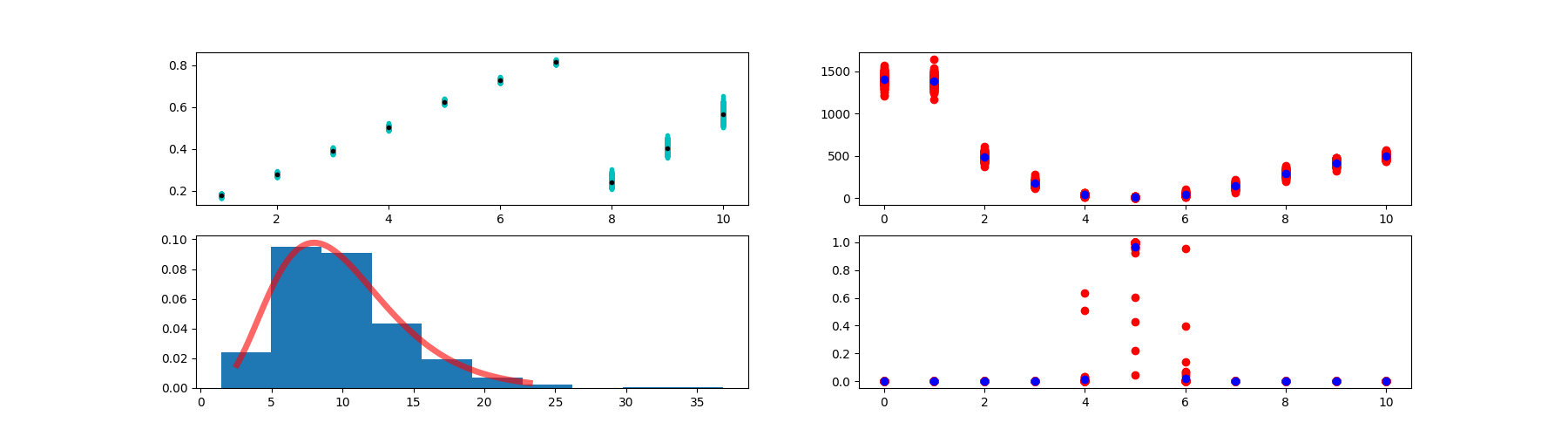}}\\
 \subfloat[distribution of particle sizes]{
 \includegraphics[trim=125 20 125 30, clip, width=.7\textwidth,draft=false]{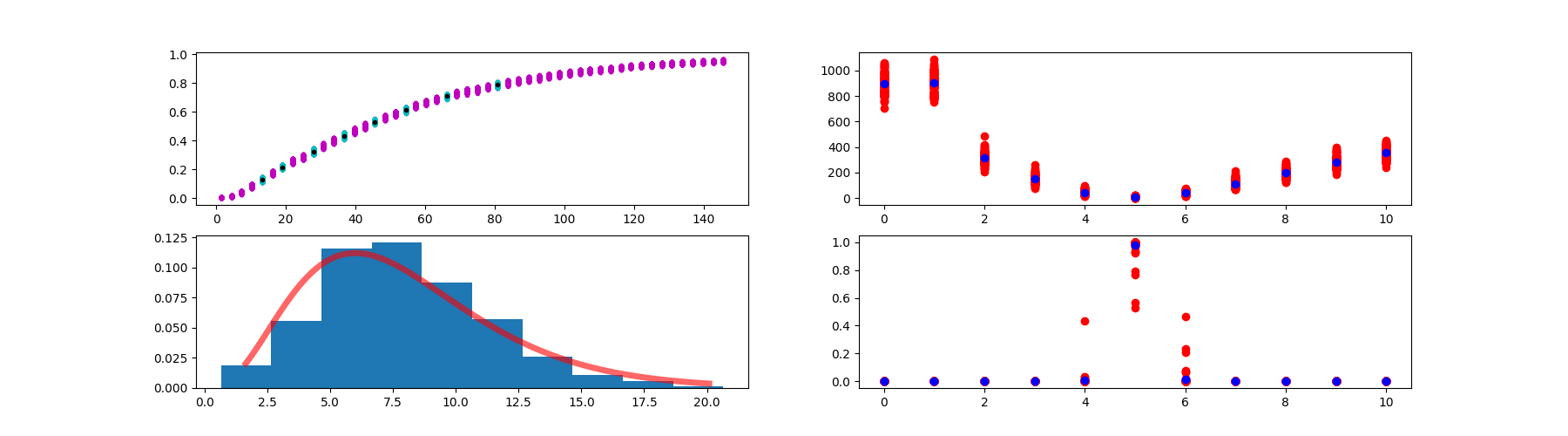}}\\
 \subfloat[$L^1$-distance + both characteristics]{
 \includegraphics[trim=125 20 125 30, clip, width=.7\textwidth,draft=false]{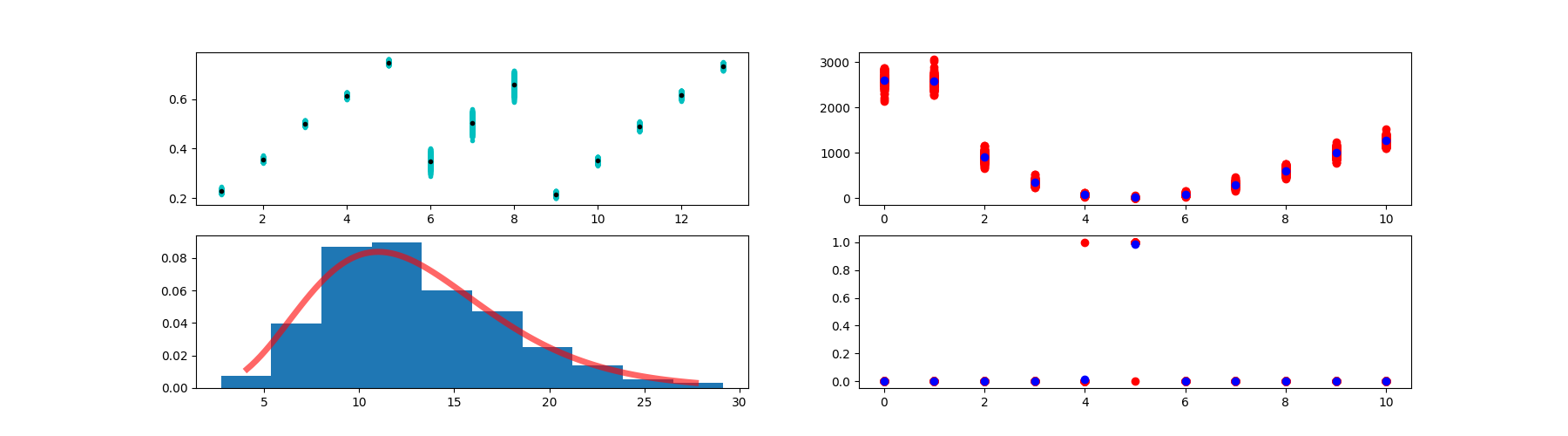}}
 \captionsetup{width=\linewidth}
 \caption{Construction of the statistical likelihood for multiple characteristics in parameter estimation method, and its evaluation for different values of jump parameter $\sigma$. The layout of sub-figures is identical to Figure~\ref{FIG:base_results}.}\label{FIG:improvements}
\end{figure}
\subsection{Multiple features in parameter estimation method}\label{SEC:multi_feature}
In this section, we again consider the basic experimental setup and discuss how the parameter identification procedure can be improved by considering additional features typical for characterizing structures of CA models as outlined in Section~\ref{SEC:Parameter_estimation}. Computing the distance between pattern data using $L^1$-norm allowed us to correctly identify the correct value of jump parameter $\sigma$ (see Figure~\ref{FIG:improvements} \corr{(a)}). However, the minimum cost is not always well pronounced. This may result in higher uncertainty in parameter identification and can be seen from the wide spread of the red dots. We can, however, significantly reduce the uncertainty by additionally using the characteristics introduced in Section~\ref{SEC:Parameter_estimation}.

The first feature we use here is the average particle size. This means that instead of computing the $L^1$-distance between two patterns, we use the absolute difference of the respective average particle sizes. We observe that if we replace the $L^1$-distance with this new mapping, the variability of the cost function is nicely reduced, and the minimum becomes more pronounced  (see Figure~\ref{FIG:improvements} (B)). An even better effect, however, can be achieved by combining these characteristics with the previously used $L^1$-distance (see Figure~\ref{FIG:improvements} (C)). Since the concatenation of the respective eCDF vectors is again Gaussian, the same approach can be applied directly. 

Next, we consider a different mapping from a pair of subsets of pattern data to one eCDF vector. That is, we do not use a distance that produces a scalar from a pair of patterns but directly create a distribution that holds the information of several scalars. Here, we employ the fact that an eCDF vector is directly available in the form of the particle size distribution of each CA pattern. Thus, we pool the particle sizes from each pair of pattern subsets (each containing $N$ patterns) and construct one eCDF vector of all the emerging scalar values. This gives us a large amount of scalar data, which results in a smooth eCDF curve. Again, all the pairs yield $n(n-1)/2$ eCDF curves. This gives us another separate feature for the parameter estimation  (see Figure~\ref{FIG:improvements} (D)). Naturally, the best results are obtained when all those features are combined. This case is shown in Figure~\ref{FIG:improvements} (E). We mainly observe the minimal variability of the cost function.

\begin{figure}\centering
\subfloat[$\sigma = 3$]{
 \includegraphics[trim=125 20 125 30, clip, width=.7\textwidth,draft=false]{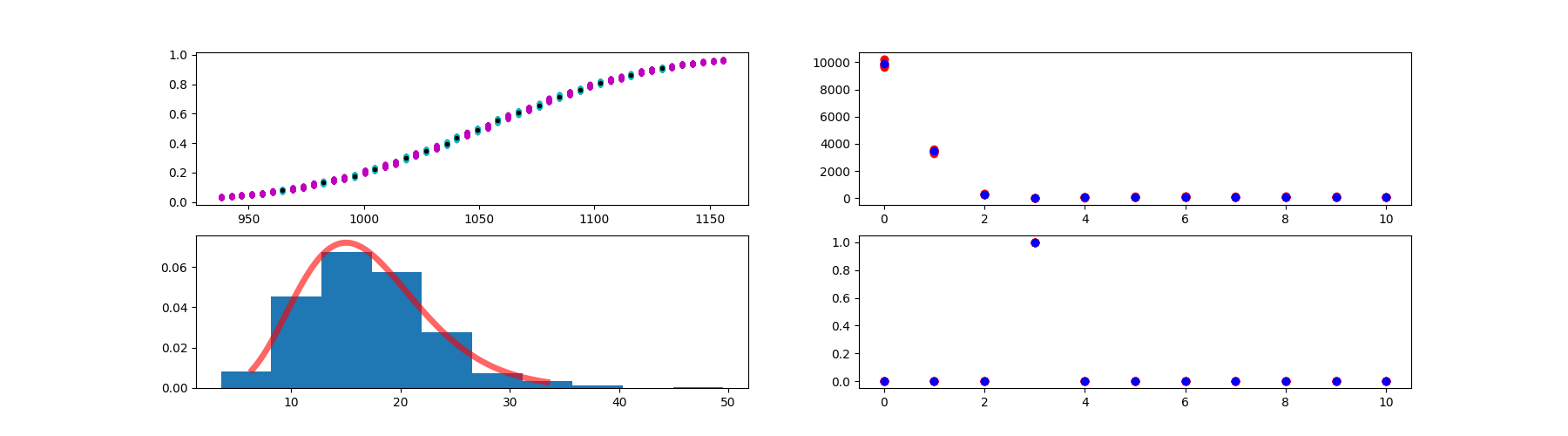}}\\
 \subfloat[$\sigma = 7$]{
 \includegraphics[trim=125 20 125 30, clip, width=.7\textwidth,draft=false]{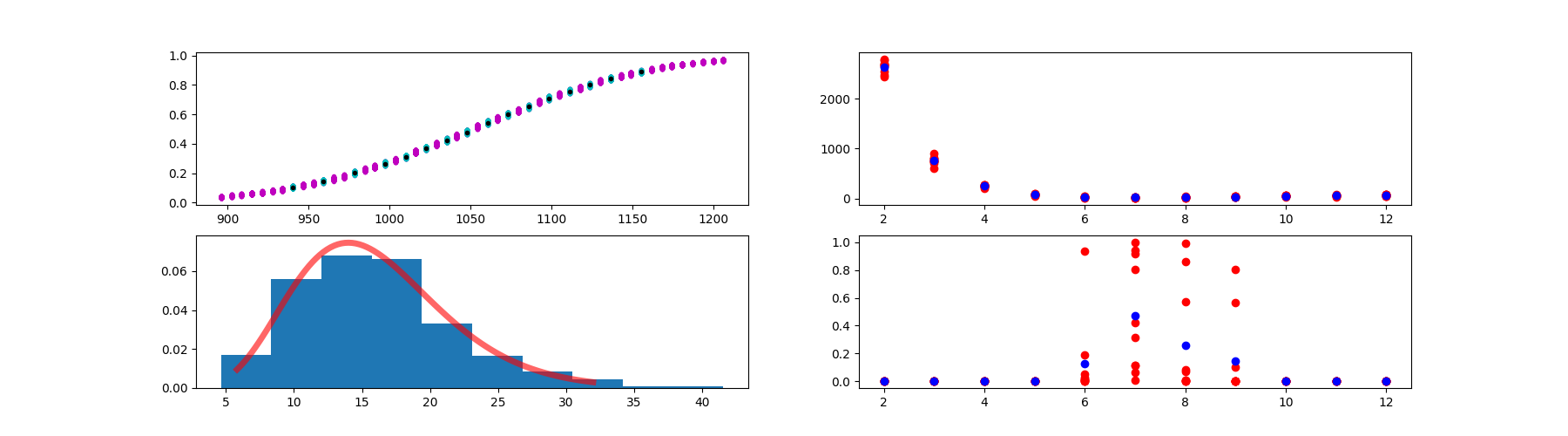}}\\
 \subfloat[$\sigma = 11$]{
 \includegraphics[trim=125 20 125 30, clip, width=.7\textwidth,draft=false]{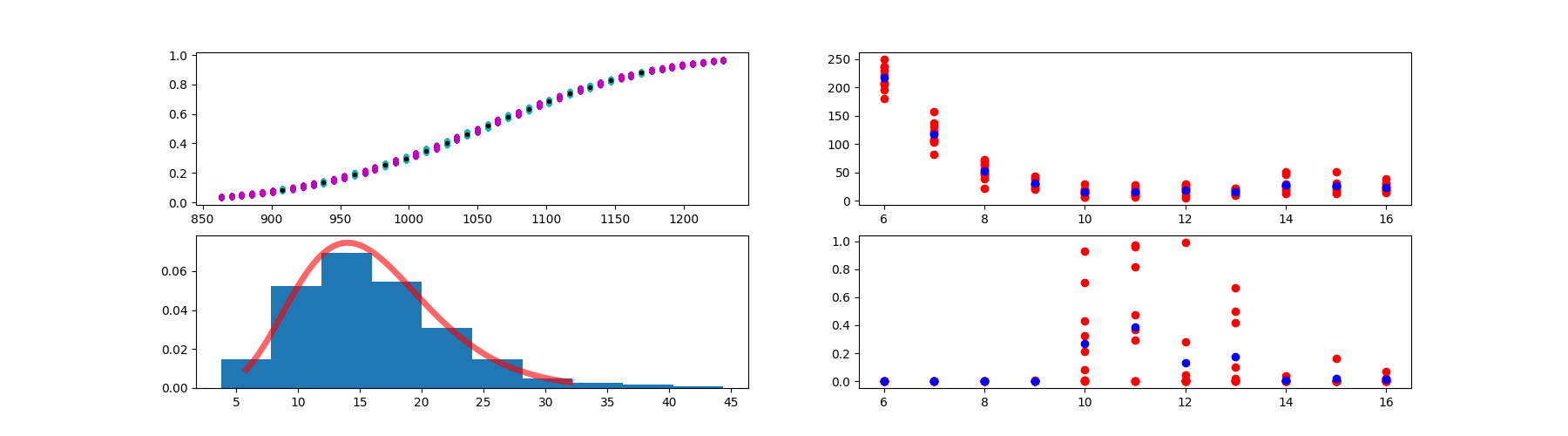}}\\
 \subfloat[$\sigma = 11$ with larger sample size]{
 \includegraphics[trim=125 20 125 30, clip, width=.7\textwidth,draft=false]{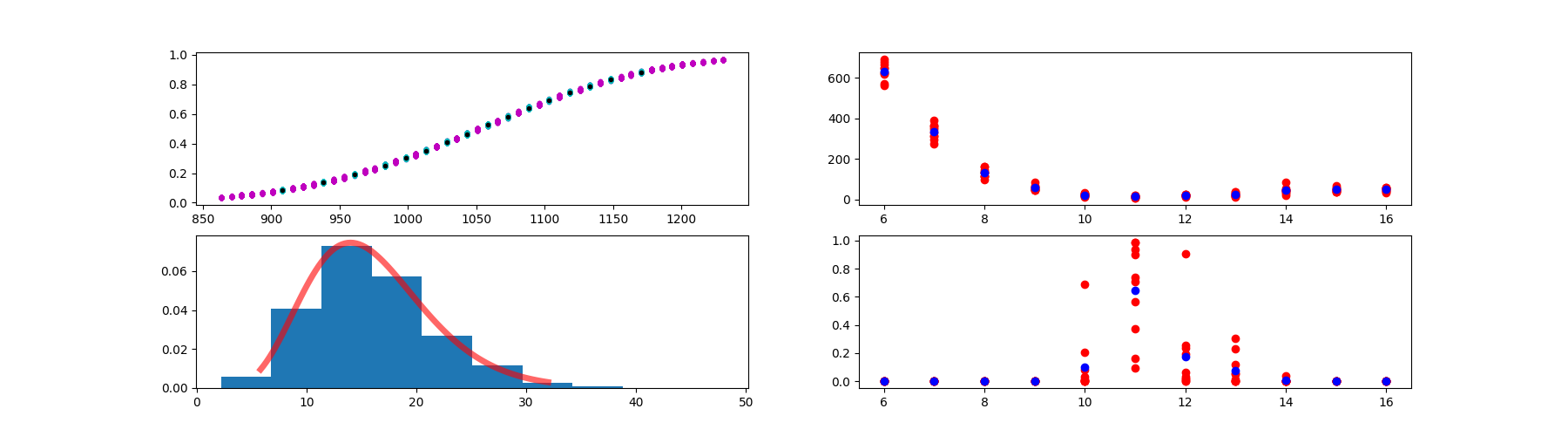}}\\
 \subfloat[$\sigma = 15$ for domain of size 25 $\times$ 25]{
 \includegraphics[trim=125 20 125 30, clip, width=.7\textwidth,draft=false]{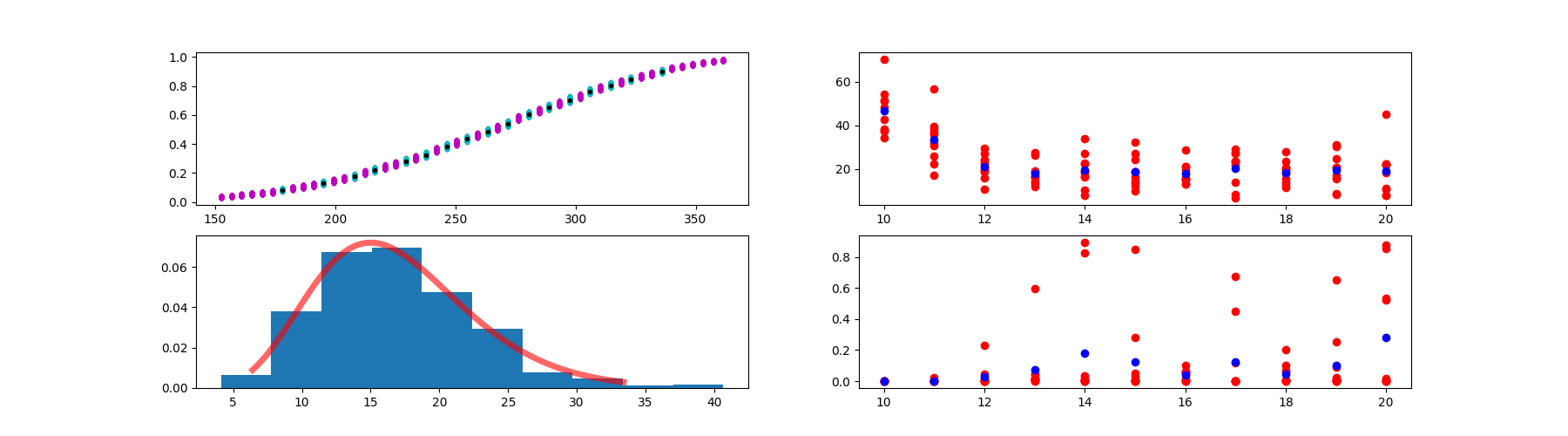}}
 % \subfloat[$L^1$-distance + both characteristics]{
 % \includegraphics[trim=125 20 125 30, clip, width=.7\textwidth,draft=false]{pictures/param_results/multiple_3.png}}
 \captionsetup{width=\linewidth}
 \caption{\corr{Construction of the statistical likelihood for multiple jump parameters $\sigma$ and an increased sample size in (d). Panel (e) demonstrates that $\sigma$ cannot be identified if it is too large compared to the domain size. The layout of sub-figures is identical to Figure~\ref{FIG:base_results}.}} \label{FIG:sigma_impact}
\end{figure}
\subsection{\corr{The impact of $\sigma$}}
\corr{Finally, we discuss the impact of the parameter that we want to identify itself. To this end, we consider $\sigma \in \{3, 7, 11\}$ for our base setup of Section \ref{SEC:results_base} in Figure \ref{FIG:sigma_impact} (a) -- (c). Our method always identifies $\sigma$ correctly but becomes less confident as $\sigma$ increases as the value for the correct $\sigma$ decreases in the bottom right panels of (a) to (c). This effect stems from our CA's property: patterns obtained for larger $\sigma$ become increasingly similar (cf.\ Figure \ref{fig:domain_porosity_jumpParameter}). This can be compensated to some extent by increasing the sample size, illustrated in Figure \ref{FIG:sigma_impact} (d). However, as $\sigma$ increases, it cannot be identified at some point since its precise value no longer influences the simulation. This stems from the spatial domain's periodicity; in this case, any particle can jump to any location independent of the precise value of $\sigma$.  Thus, our method works for different $\sigma$, but the sample sizes needed to obtain confident parameter estimates vary depending on the actual values of $\sigma$. However, the limitation of identifying too large $\sigma$ as compared to the domain size remains and cannot be cured by using more data since changes in $\sigma$ do have no effect if $\sigma$ is chosen such that all particles can jump anywhere in the domain. This effect is illuminated in Figure \ref{FIG:sigma_impact}'s panel (e), demonstrating that our method correctly excludes the possibility of $\sigma < 12$ and correctly shows that large enough sigma values (here $\sigma=15)$ can not be identified (by any method)  since other (large enough) $\sigma$ produce the same patterns.
}
\section{Conclusions}
In this work, we introduced a parameter identification that works well for discrete problems, as in the context of CA models. Our parameter estimation method allowed us to identify model parameters using pattern data only, without knowing about initial states. We demonstrated the method's applicability by successfully identifying the value of jump parameter $\sigma$ from a set of $4000$ patterns. Moreover, we proved the robustness of our approach for different configurations of the model for the number of selected bins, the domain size, and the number of CA steps. Finally, we showed that the accuracy could be drastically improved if commonly used CA pattern data features were incorporated into our method's scheme.

\corr{Possible limitations of our approach stem from two main sources: CAs as base models and the statistical origin of the method. The main drawback of CAs is the loose physical motivation of their rules, making the validation of model predictions against experimental data more challenging. Especially when the complexity of the physical process being modeled by CA grows, the problem of ruling out the wrong model mechanisms becomes more challenging.}

\corr{
The second potential limitation of our approach stems from its statistical formulation. Indeed, a single evaluation of the statistical cost function $f_{\bm s_\textup{data}} (\bm{\sigma})$ requires computing repeated simulations of the underlying model. The numerical code we use in this work can very efficiently compute a single simulation of the CA. This means that the computational time needed for evaluating the $N$ model grows linearly concerning the number of simulations. While it is not a problem for the situation where the cost of a single model run is low, it might become prohibitive for more complex CA models, which require more time to run. One possible solution to this problem could be the development of solvers optimized for batched simulations, as was done in \cite{KazarnikovH20, KazarnikovRHLR23}.
}

Future work may include applying the parameter estimation method to more than one parameter as already outlined in \cite{haario2015} and combined with various characteristics/norms. Additionally, more sophisticated cellular automaton models may be used. This approach could address realistic problems, improve understanding of the underlying process, and draw conclusions relevant to real-life problems.
\section*{Statements \& Declarations}
% 
% We kindly acknowledge the fruitful discussions with Alexander Prechtel and Simon Zech on cellular automaton methods with applications to soil science.
% % 
% \subsection*{Funding}
% % 
% A.\ Rupp has been supported by the Academy of Finland's grant number 350101 \emph{Mathematical models and numerical methods for water management in soils}, grant number 354489 \emph{Uncertainty quantification for PDEs on hypergraphs}, grant number 359633 \emph{Localized orthogonal decomposition for high-order, hybrid finite elements}, Business Finland's project number 539/31/2023 \emph{3D-Cure: 3D printing for personalized medicine and customized drug delivery}, and the Finnish \emph{Flagship of advanced mathematics for sensing, imaging and modeling}, decision number 358944. This research has also been supported the German research foundation's research unit 2179 \emph{MAD Soil}, and the DAAD PPP Finnland under grant number 57610378.
% 
\subsection*{Competing Interests}
The authors have no relevant financial or non-financial interests to disclose.
% 
% \subsection*{Author Contributions}
% % 
% All authors contributed to the study's conception and design. The simulation and parameter estimation software was designed by Andreas Rupp and implemented by Andreas Rupp, Alexey Kazarnikov, and Joona Lappalainen. Heikki Haario and Alexey Kazarnikov conceptualized the eCDF method, and Nadja Ray contributed the characteristics. Nadja Ray and Andreas Rupp wrote the first draft of the manuscript, and all authors commented on previous versions. All authors read and approved the final manuscript.
% 
\subsection*{Data Availability}
All used datasets can be created using the codes in \cite{RuppZL22,RuppHK22}, freely available software packages.

\bibliographystyle{ARalpha}
\bibliography{CAMparameters}
\end{document}